\RequirePackage{fix-cm}

\documentclass[twocolumn]{svjour3}          
\smartqed  
\usepackage{graphicx}
\usepackage{url}
\usepackage{algorithm}
\usepackage[noend]{algpseudocode}
\usepackage[caption = false]{subfig}
\usepackage{booktabs,makecell}
\usepackage{picins}
\usepackage{mwe,wrapfig}

\begin{document}

\title{Anomaly Detection and Modeling in 802.11 Wireless Networks}


\author{Anisa Allahdadi         \and
        Ricardo Morla 
}


\institute{A. Allahdadi \at
              INESC TEC, Faculty of Engineering, University of Porto, \\
              Campus da FEUP, Rua Dr. Roberto Frias N8 378, \\
              4200-465 Porto, Portugal \\
              \email{anisa.allahdadi@inescporto.pt}          
           \and
           R. Morla \at
              \email{ricardo.morla@fe.up.pt}
}


\maketitle

\begin{abstract}

IEEE 802.11 Wireless Networks are getting more and more popular at university campuses, enterprises, shopping centers, airports and in so many other public places, providing Internet access to a large crowd openly and quickly. The wireless users are also getting more dependent on WiFi technology and therefore demanding more reliability and higher performance for this vital technology. However, due to unstable radio conditions, faulty equipment, and dynamic user behavior among other reasons, there are always unpredictable performance problems in a wireless covered area. Detection and prediction of such problems is of great significance to network managers if they are to alleviate the connectivity issues of the mobile users and provide a higher quality wireless service. This paper aims to improve the management of the 802.11 wireless networks by characterizing and modeling wireless usage patterns in a set of anomalous scenarios that can occur in such networks. We apply time-invariant (Gaussian Mixture Models) and time-variant (Hidden Markov Models) modeling approaches to a dataset generated from a large production network and describe how we use these models for anomaly detection.  We then generate several common anomalies on a Testbed network and evaluate the proposed anomaly detection methodologies in a controlled environment. The experimental results of the Testbed show that HMM outperforms GMM and yields a higher anomaly detection ratio and a lower false alarm rate. 


\end{abstract}

\section{Introduction}
\label{intro}

Wireless 802.11 networks are getting more and more popular in providing Internet access for a large number of users in university campuses, enterprises, urban areas, and many other public places. These large-scale networks, particularly speaking of IEEE 802.11 Infrastructure mode, consist of basic network components: Wireless Stations, wired stations, and the Access Points (AP) that function as connection links between the wired and wireless sections. The APs provide coverage and capacity for supporting mobile clients with heterogeneous devices and a variety of applications. Among the many characteristics of such large-scale network is the transition of huge volumes of traffic as a result of intensive usage from different locations all over the coverage area. The mobile clients demand reliable connection and high performance in all circumstances and expect their applications to work smoothly around the wireless covered field, but this is an ideal case which is not always achievable. The wireless users, most of the time suffer from low coverage, intermittent connectivity, authentication failure, degraded performance and many other complications originated from the unreliable nature of wireless connection and dynamic usage pattern of other users in the vicinity.  

Having further explored the connectivity procedure in wireless networks, some inherent concerns and dilemmas become more clear. In Wireless 802.11 networks, mobile stations perform an active or passive scanning process to discover available APs in the vicinity and connect to an AP with the highest received signal strength (RSS) \cite{Ref0}. This association strategy, only based on RSS, can lead to many connectivity problems and performance issues as it may result in significant load imbalance between APs. The overloaded APs can still present high RSS and try to accommodate more stations while other APs are only slightly loaded or even idle. Another source of performance degradation in WLANs is the multi-rate flexibility and the fairness mechanism of the MAC protocol- when a station far from the AP reduces its bit rate to avoid repeated unsuccessful frame transmission and as a result degrades the throughput of the other stations associated with the same AP \cite{Ref1}. In addition to the aforementioned problems, due to the unreliable and time-varying nature of the wireless channels, 802.11 networks usually suffer from many pitfalls such as exposed and hidden terminals, capture effect, interferences, signal fading, inconsistent coverage, and many other examples. In such circumstances, high packet loss is observed \cite{Ref2}- that results in inconsistent connectivity and low performance. Network managers are concerned about discovering such sort of problems and abnormal events that occur in their network. Detection of anomalies is not only advantageous for prompting immediate administrative actions but also useful for long-term network design, planning, and maintenance decisions as the network infrastructure and usage evolve over time.

In large deployments of 802.11 networks with varying usage, channel conditions, and operational constraints, network managers often demand tools that provide them with a comprehensive view of the entire network for automatic detection of the problems. In such widespread networks, where at any moment there is a high possibility of mal-functioning of APs and user devices, the necessity of such automatic tools or applications is vital to preserve the quality of service at an acceptable level. Monitoring the infrastructure by any means rather than intelligent diagnostic tools seems inconvenient in practice or overpriced in budget. For example, it is expensive to deploy third party devices like sensors and sniffers individually on clients machines or APs for detection of problems in different OSI layers, as studied earlier \cite{Ref5,Ref6,Ref7}. And it seems impractical for network staff to walk around the wireless covered area with a device in their hand monitoring the network and measuring the quality of connections at any time. In this paper, we propose an automatic diagnostic tool that analyzes the usage data of the APs- collected from a RADIUS authentication server. We apply probabilistic learning algorithms to produce a model for each access point or group of access points, and identify anomalous events with a margin of certitude. AP usage modeling and anomaly detection in hotspots would assist network administrators to ensure long-term quality of service by analyzing various connectivity factors of wireless users in particular localities. 
We propose probabilistic graphical model- and in particular HMM- to establish a comprehensive image of the evolving structure of wireless networks, to distinguish usage behaviors in different locations and grouping context and their correlations and dependencies, and to represent the spatio-temporal anomalous patterns detected in wireless networks. In the current work we focus more on proposing individual models for APs as the ground truth data is only available through the single AP Testbed deployment and the multiple APs' experiment is planned for the future work. The prospective methodology is based on the development of HMM models and a detection tool using WiFi campus data; our recent contributions \cite{Ref23,Ref24} have taken this approach into account. As a preliminary investigation on the subject, we focused on short 802.11 sessions recorded through RADIUS authentication as a network artifact and an indicator of quality of wireless access \cite{Ref23}. In \cite{Ref24} an exhaustive analysis is performed for outlier detection in 802.11 wireless networks using HMM variations- single HMM, mixture of HMMs and individual HMMs- and is evaluated by the state of the art statistical methodologies. Furthermore a number of network anomalous patterns are represented, in the same study, considering HMM parameters such as hidden states' transition and partial likelihood of the observation sequences. In the present study we considered HMM and its counterpart time-invariant methodology- Gaussian Mixture Model (GMM)- to investigate the temporal relevancy of the employed data, whether a simpler time-invariant model such GMM is adequate to detect anomalies or a more complex model like HMM is really needed. These two methodologies are analyzed and compared with each other both in modeling and anomaly detection experiments. 


This paper contains two main parts: 1) analysis and modeling of 802.11 AP usage and exploring the time dependency of the employed data, and 2) identification and detection of different types of anomalies and characterizing them efficiently. The aforementioned objectives are investigated on a large dataset of AP usage and examined on a smaller scale testbed for the purpose of evaluation.    

The rest of the paper proceeds as follows. In section 2, the related work and the most recent researches relevant to the current work are presented. In section 3, the wireless setup procedure in infrastructure mode is characterized and the key attributes and functionalities of RADIUS protocol are defined. Section 4 deals with the process of data accumulation as a result of wireless users' association attempts, and presents a set of main features extracted from the dataset and feature selection techniques for further analysis. Statistical modeling of the AP usage data, categorized as time-invariant and time-variant approaches are provided in section 5 and a brief discussion is enclosed at the end of the section comparing these methodologies. In section 6, we describe how the proposed models serve to detect and characterize different forms of anomalies. In section 7, the experimental results are analyzed and discussed, and the evaluation process is represented based on the deployed testbed in a controlled environment. In section 8, the major conclusions are provided and prominent directions for future work are identified. 

\section{Related Work}
\label{rel-work}

\subsection{Wireless Measurement Tools}
Several prior works are dedicated to studying the dynamics of wireless network behavior, as well as the performance and reliability of WLAN technologies \cite{Ref6} \cite{Ref10} \cite{Ref11} \cite{Ref12}. In \cite{Ref6} a system called Jigsaw is presented which uses multiple monitors to provide a single unified view of physical, link, network and transport-layer activities, including inference techniques for the particular issues of 802.11. The authors deployed an infrastructure with over 150 radio monitors that capture 802.11b and 802.11g activities in a university building to investigate the causes of performance degradation. Significant challenges of such vast distributed monitoring system include the necessity of hardware and software instrumentations on each and every monitor and the scalable synchronization difficulties and inaccuracies. For this reason, most wireless management techniques avoid broad modifications in the clients devices, sensors, sniffers and monitors deployed in the large wireless covered area. 

In another line of research a Passive Interference Estimator (PIE) is presented \cite{Ref10} which provides a fine-grained estimation of link interferences in WLAN. PIE provides an estimate of WLAN interference caused by client mobility, dynamic traffic loads, and varying channel conditions. This work is inspired by two previous WLAN monitoring approaches: the aforementioned Jigsaw \cite{Ref6} and WIT \cite{Ref19}. The PIE producers use sniffing at APs to avoid deploying additional monitors similar to Jigsaw, but with the penalty of missing a portion of uplink client traffics and hence uplink client conflicts. However, they proposed an accurate approach in estimating link interference by providing a conflict graph in real time. 

In a similar direction of work, fine-grained detection algorithms are proposed that are able to distinguish the root-causes of performance degradation at the physical layer \cite{Ref11}. It is described that various faults, such as hidden terminals, capture effects and noise, could have the same propagation effects on the network layer (degraded throughput) and therefore could lead to the same remediation techniques from 802.11 (rate fallback), while they have completely different origins in the physical layer. Hence, the researchers of this work designed a unified framework for this purpose, called MOJO, that combines the observations from multiple distributed sniffers and diagnoses the granularity of the root causes to suggest appropriate remedies for different physical faults. Although the proposed framework measures the impact of the most commonly observed faults on different network layers, it is still a client side monitoring system and suffers from the extensive sniffer distribution all over the wireless covered area.

In \cite{Ref12}, WiMed is proposed that uses only local measurements from commodity 802.11 NICs for understanding how the medium is utilized, and for inspecting the causes of interferences (including non-802.11 devices). WiMed provides a time-domain view of how the medium is used in a given 802.11 channel, and identify the root causes of interference using physical layer properties such as bit error patterns and medium busy times. The authors refrain from elaborate instrumentation and dedicated infrastructure, however detectors are only implemented for interference and contention, and there is a higher confidence for recognition of non-802.11 interferer rather than 802.11 sources of interference.   

All the above literatures expose the difficulties in monitoring the wireless environment thoroughly, and the challenges of performance estimation in such complex networks. Most cases- require heavy instrumentation of the user devices and focus on specific anomalies affecting individual users- thus neither considering usage trend nor location related anomalies.

\subsection{Usage Modeling and Anomaly Detection}
There are several lines of research that take an approach closely related to our work. In \cite{Ref13} AP usage and daily keep-alive events of mobile stations in 802.11 hotspots in infrastructure mode are analyzed and modeled. In this work, generative probabilistic models are investigated such as Gamma mixture of exponentials and Conditional probability models considering dependencies between consecutive samples in time. The generative statistical models and experimental results of this work – conducted on a very similar dataset to ours – provide some broad insight into AP usage and illustrate those aspects of such networks that benefit our work. 

In \cite{Ref14}, a usage pattern called "abrupt ending" is explored in a similar dataset, and it concerns the disassociation of a large number of wireless sessions in the same AP within a one second window, or in a nutshell "simultaneous session ending". The authors of this work, further investigate this concept and introduce some anomalous patterns that might be in correlation with the occurrence of this phenomena. For instance, they propose that interference across the AP vicinity could be deduced when abrupt endings happen to neighboring APs within specified time interval, or the AP overloaded could be inferred when the continued sessions are present after abrupt endings. There are a number of other anomalous patterns reported in this paper such as AP halt/crash, persistence interference and intermittent connectivity. The classification and analysis of these anomaly-related patterns performed in this research, inspired our work to regenerate similar anomalies in a real Testbed to experiment and evaluate the HMM methodologies practiced in the current study. 

\subsection{HMM Applications in Network Analysis}
In wireless networking, HMMs are employed to address various aspects of network measurement and analysis. Hierarchical and Hidden Markov based techniques are analyzed in \cite{Ref15} to model 802.11b MAC-to-MAC channel behavior in terms of bit error and packet loss. The authors employed two random variables in packet loss process, inter-arrival-rate and burst-length of packet loss, and applied the traditional two-state Markov chain. The results demonstrates that two-state Markov chain provides an adequate model for the 802.11b MAC-to-MAC packet loss process. Furthermore, in regard to bit error modeling, three other Markov-based chains are evaluated: full-state, hidden, and hierarchical Markov chains. Among these chains, it is illustrated that the full-state Markov bit error model of order 9 and above, renders the best performance. Since the main concern to use HMM in this example is to generate error traces, a simple three-state HMM is designed and utilized for one HMM solution: the adjustment of model parameters to best account for the observed signal.   

In a more recent line of research in \cite{Ref16} a multilevel approach involving HMMs and Mixtures of Multivariate Bernoullis (MMB) is proposed to model the long and short time scale behavior of wireless sensor network links, that is, the binary sequence or trace of packet receptions (1s) and losses (0s) in the link. In this approach, HMM is applied to model the long-term evolution of the trace, and the hidden states correspond to packet reception rate. Within the aforementioned hidden states, the short-term evolution of the trace is modeled by either another HMM or by a MMb. That is how the multilevel, or in this case the two level approach, is formed. The notion of multilevel HMM, or higher dimensional HMM, is an impressive concept regarding to our own work, and we intend to make use of this approach to improve our HMM variations for anomalous pattern detection in the future work.  

One of the salient applications of HMMs addressed in wireless networking is prediction. For instance in \cite{Ref17} HMMs are utilized to model and predict the spectrum occupancy of sharing radio bands. The channel status prediction is considered as a binary series prediction problem, as channel occupancy can be represented as idle or busy depending on the presence or absence of a primary user activity. An ergodic two-state discrete HMM deals with this problem. Some other prominent work has been done on a very similar subject in radio spectrum sensing and status prediction using HMMs in \cite{Ref20,Ref21,Ref22}.

Furthermore, in another related work, HMMs are applied for modeling and prediction of user movement in wireless networks to address issues in Quality of Service (QoS) \cite{Ref18}. User movement from an AP to an adjacent AP is modeled using a second-order HMM. Although the authors demonstrated the necessity of using HMM instead of Markov chain model, the proposed model is only practical for small wireless networks with a few number of APs, not huge enterprises or widespread campuses.

As the above literatures indicate and to the best of our knowledge, HMM related studies in wireless network management are rarely used specifically in performance anomaly detection.

\section{Wireless Setup in Infrastructure Mode}
\label{wrls setup}

In this section we describe how a 802.11 station associates to an access point and how our setup authenticates the user and authorizes access to the network.

\subsection{Association of Wireless Station to Access Point}
\label{assoc}

\begin{table*}[!htbp]
\centering
\caption{The Key Attributes of RADIUS Accounting Table}
\label{tab:1}       
\begin{tabular}{| p{3.0cm} | p{12.5cm} |}
\hline
Acct-Status-Type & has three values: Start, Alive and Stop. A Start record is created when a user session begins. An Alive record is registered after each 10 or 15 minutes for the users that are still connected. A Stop record is generated when the session ends. \\ \hline
Acct-Session-Id & is a unique number assigned to each session to facilitate matching the Start and Stop records in a detail file, and to eliminate duplicate records. \\ \hline
Acct-Session-Time & records the user's connection time in seconds. This information could be included in Alive or Stop records. \\ \hline
Acct-Delay-Time & is the number of seconds passed between the event and the current attempt to send the record. The approximate time of an event can be determined by subtracting the Acct-Delay-Time from the time of the record's arrival on the RADIUS accounting server. \\ \hline
Called-Station-Id \& Calling-Station-Id & record the IP address of the AP (Called Station) and the wireless user (Calling Station) connected to that AP. \\ \hline
Timestamp & records the time of arrival on the RADIUS Accounting host measured in seconds since the epoch (00:00 January 1, 1970). It provides a machine-friendly version of the logging time at the beginning of the accounting record. \\ \hline
Acct-Input-Octets \& Acct-Output-Octets & records the number of bytes received (Acct-Input-Octets) and sent (Acct-Output-Octets) during a session. These values appear in Alive or Stop records. \\ \hline
Acct-Input-Packets \& Acct-Output-Packets & records the number of packets received (Acct-Input-Packets) and sent (Acct-Output-Packets) during a session. These values appear in Alive or Stop records. \\ \hline
\end{tabular}
\end{table*}


The process of the association of a wireless mobile station to an AP, as it is currently implemented by most manufacturers is described as follows:
A wireless station scans the available channels of each AP in the neighborhood and listens to the beacon (passive approach) or probe response frames (active approach). IEEE 802.11 protocol defines a number of Wi-Fi channels ranging from 2.4 GHz to 5.9 GHz. The Wi-Fi channels that are the concern of this work (802.11 b/g/n) are listed in the 2.4 GHz range and consist of one to eleven channels (up to fourteen in some countries). The wireless station stores the received signal strength indicator (RSSI) of the APs in the vicinity and other relevant information such as extended service set identification (ESSID), encryption type (e.g. WPA, WEP), etc. When the scanning process is over, the wireless station selects an AP with the highest RSSI among the observed APs in its proximity. After the process of authentication/ authorization is accomplished, the permission is granted to the wireless station and the connection is established. Forthwith, the wireless station is associated with the new AP and the user is ready to send and receive traffic through that AP. The wireless station will be disassociated from the current AP under the mobility circumstances, AP shutdown or halt, RSSI recession or some other normal or abnormal consequences of network fluctuations. The process of AP selection only based on the strongest RSSID lead to aforesaid load imbalance problem, while some APs are overcrowded and the other available APs remain idle.

\subsection{Remote Authentication Dial-In User Service (RADIUS)}
\label{radius}

Remote Authentication Dial-In User Service (RADIUS) is a network protocol that enables remote access servers to communicate with a central server to authenticate dial-in users and authorize their access to the requested system or service. RADIUS is commonly used by Internet Service Providers (ISPs), cellular network providers, and corporate and educational networks, and it allows the management of user profiles in a central database that all remote servers can share. Having a central service facilitates the process of tracking usage for billing and network statistics. RADIUS is a de facto industry standard used by a number of network product companies and it is a proposed IETF standard \nocite{Ref27}. This protocol is used to provide network authentication, authorization, and accounting services, and it is particularly described in Request for Comments (RFC) 2865 \nocite{Ref25} and RFC 2866 \nocite{Ref26}.

According to RADIUS protocol, whenever a client associates to an 802.11 AP, a log event "START" is recorded in the accounting database. While the client is still connected to this AP, every 10 or 15 minutes (based on the server configuration) an interim log event "ALIVE" is issued to refresh the connection between the client and the AP. Eventually, when the user decides to disconnect from the network, or for some reason it is forced to leave the network, a log event "STOP" is recorded, which marks the end of the association period of this user. Each log record includes some key attributes of time-stamp, session ID, association duration, number of input and output packets/octets. Table \ref{tab:1} present a brief explanation of some of these key attributes more relevant to this work.

RADIUS serves three main functionalities: 

\begin{itemize}
  \item Authenticates users before granting them access to the network.
  \item Authorizes the authenticated users for specific network services.
  \item Accounts the usage activity of the authorized users for the services in use.
\end{itemize}

AAA stands for "Authentication, Authorization, and Accounting". It defines an architecture that authenticates and grants authorization to users and accounts for their activity. When AAA is not used, the architecture is described as "open", where anyone can gain access and do anything, without any tracking.


\subsubsection{Authentication and Authorization}
\label{auth-auth}

Authentication refers to the process of validating the identity of the user by matching the credentials provided by the user on the AAA server. If the credentials match, the user is authenticated and gains access to the network. On the contrary, if the credentials mismatch, authentication fails and network access is denied. Authentication can also fail, due to user incorrectly entering the credentials. A network administrator can choose to permit limited network access to unknown users, for instance the guests of a conference or a temporary public event in academic environments. 

\noindent
\begin{figure}
\centering
\includegraphics[width=0.5\textwidth]{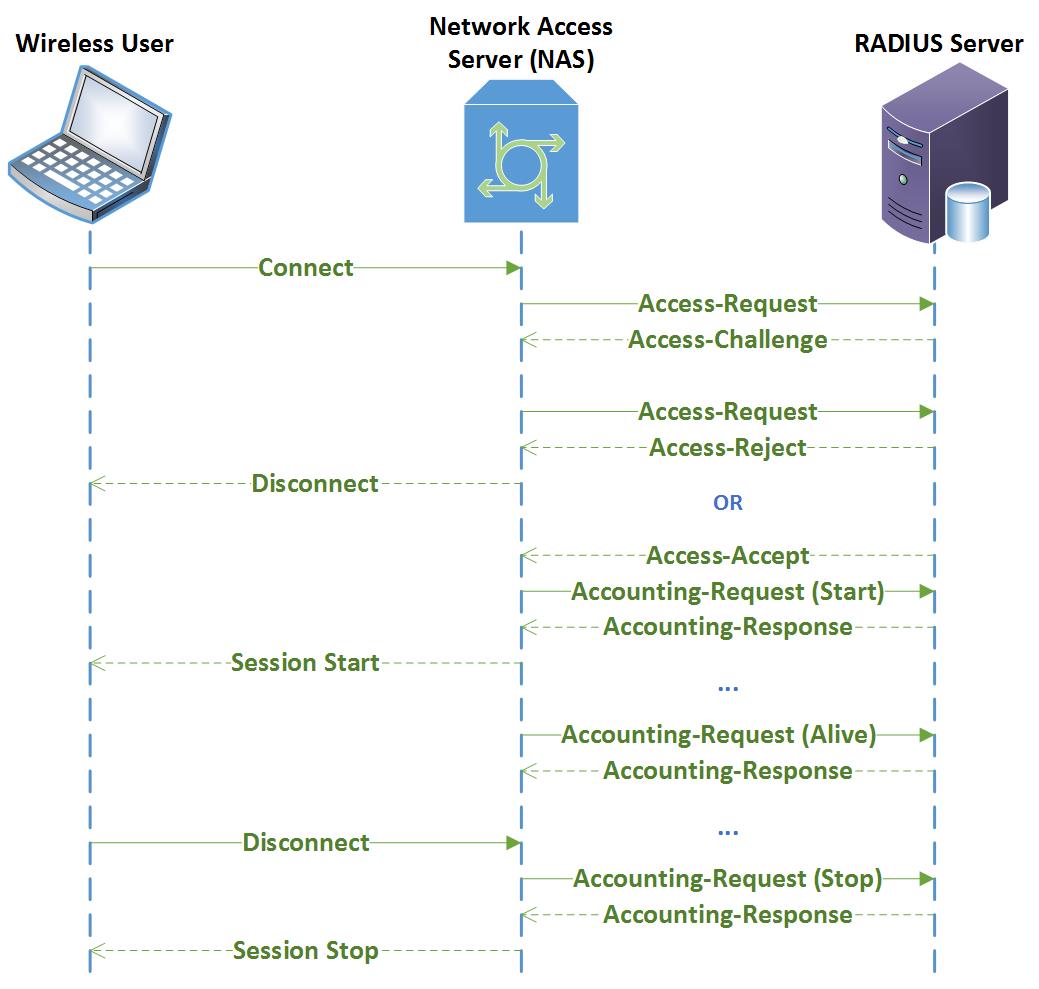}
\caption{The Authentication and Authorization Process in RADIUS}
\label{fig:40}      
\end{figure}

\begin{table*}[!htbp]
\centering
\caption{The semester-level evolution of hotspot usage during two years}
\label{tab:2}       
\begin{tabular}{| p{3.1cm} | p{1.8cm} | p{1.8cm} | p{1.8cm} | p{3.2cm} | p{3.2cm} |}
\hline
Academic Semesters & \# APs & \# Users & \# Sessions & Total Input Traffic (TB) & Total Output Traffic (TB)  \\
\hline\hline
Spring 10/11 & 238 & 15564 & 5127823 & 148 & 253 \\ \hline
Fall 10/11 & 278 & 15614 & 2619497 & 81 & 138 \\ \hline
Spring 11/12 & 317 & 20200 & 5879742 & 177 & 359 \\ \hline
Fall 11/12 & 338 & 21946 & 7167023 & 91 & 170 \\ \hline
\end{tabular}
\end{table*}

Authorization deals with the process of deciding what permissions are granted to the user. For example, the user may or may not be permitted certain kinds of network access or allowed to issue certain commands. Typically, a user login consists of a query (Access-Request) from the NAS to the RADIUS server and the RADIUS server either grants or denies authorization (Access-Accept or Access-Reject) based on the information passed by in the request query. In each case, the RADIUS server manages the authorization policy and the NAS enforces the policy. The process of authentication and authorization is delineated in Figure \ref{fig:40}.



\subsubsection{Accounting}
\label{accounting}
Accounting refers to the recording of resources users consume during the time they are connected to the network.
The information gathered can include the total system time used, and the amount of data sent or received by the user during a session. Over a network session, the NAS periodically sends an accounting data of user activity to the server (in "Alive" or "Stop" sessions). This data is mainly used for the billing purposes. However, we used the accounting information for the reason of network monitoring and management as the log dataset is already stored in a central database, the RADIUS server, and facilitates the data collection process. 

The detailed information of users' activities is not included in the summary sent by NAS- for instance the visited web sites or particular protocols in use is local to the NAS- and is not available to the RADIUS server. 
Transactions between the client and RADIUS server are authenticated through the use of a shared secret, which is never sent over the network. In addition, user passwords are sent encrypted between the client and RADIUS server to eliminate the possibility of snooping on an insecure network.


\section{Data Description and Feature Selection}
\label{data-desc}

In this section we present the main dataset used in this paper, provide some preliminary statistical analysis and describe the key features emerge from the raw dataset as well as the of process of feature selection for modeling and further investigations.
 
\noindent
\begin{figure*}
\centering
\raisebox{\dimexpr-.5\height-1em}{\includegraphics[width=0.99\textwidth,height=5.7cm]{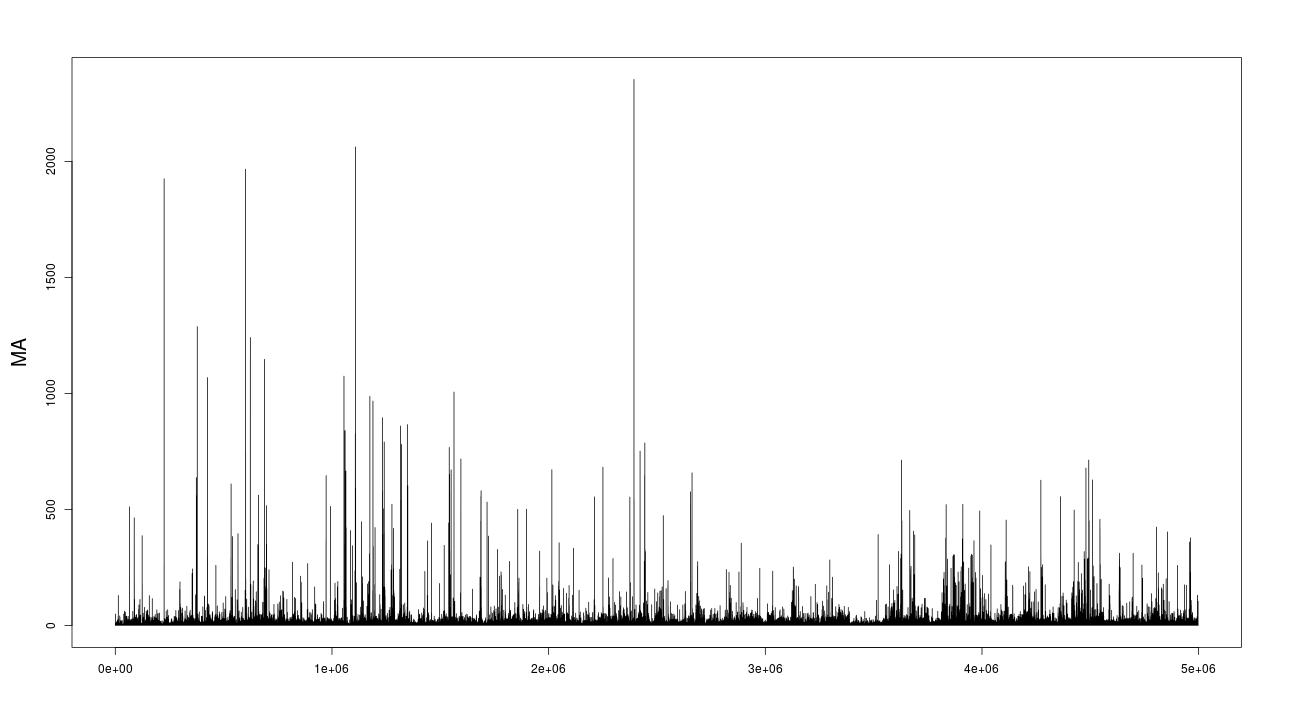}}\ \subfloat[\label{fig:4a}]{} \\[\topskip]
\raisebox{\dimexpr-.5\height-1em}{\includegraphics[width=0.99\textwidth,height=5.4cm]{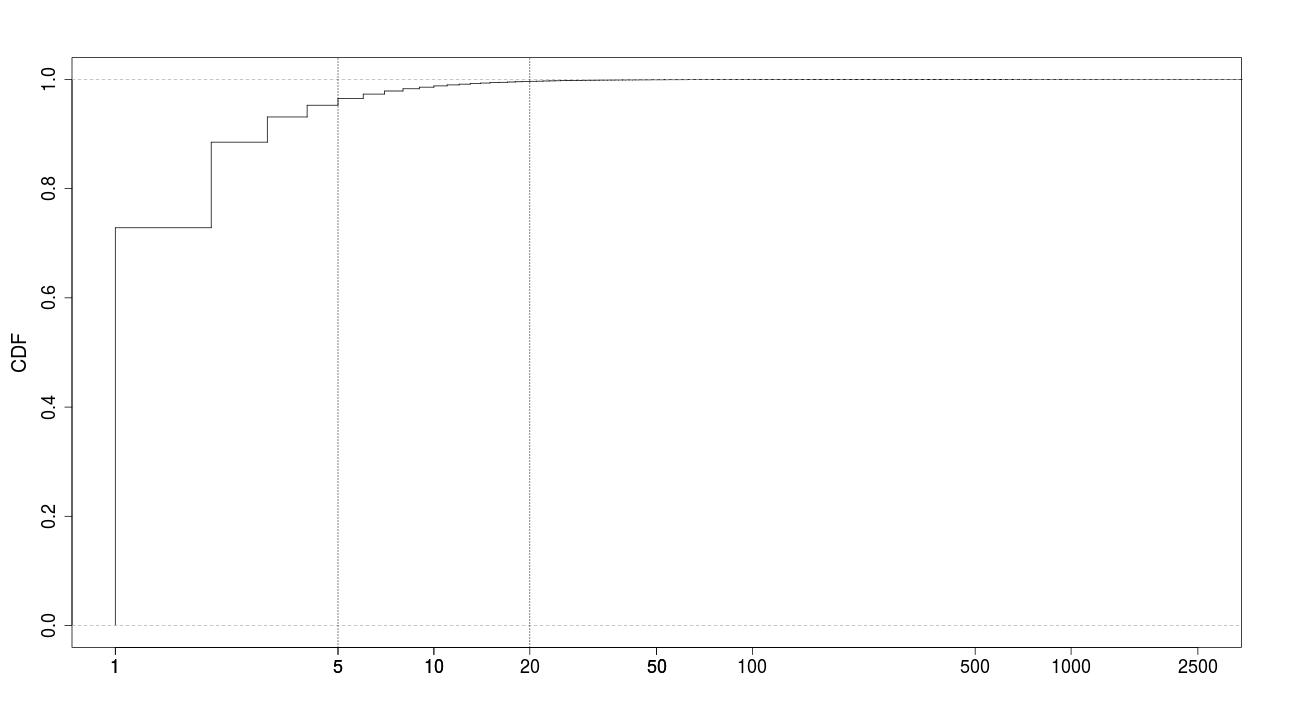}}\ \subfloat[\label{fig:4b}]{}
\caption{Number of Sessions per User (Hourly) \protect\subref{fig:4a} Moving Average, and \protect\subref{fig:4b} CDF}
\label{fig:4}      
\end{figure*}

\noindent
\begin{figure*}
\centering
\raisebox{\dimexpr-.5\height-1em}{\includegraphics[width=0.99\textwidth,height=5.7cm]{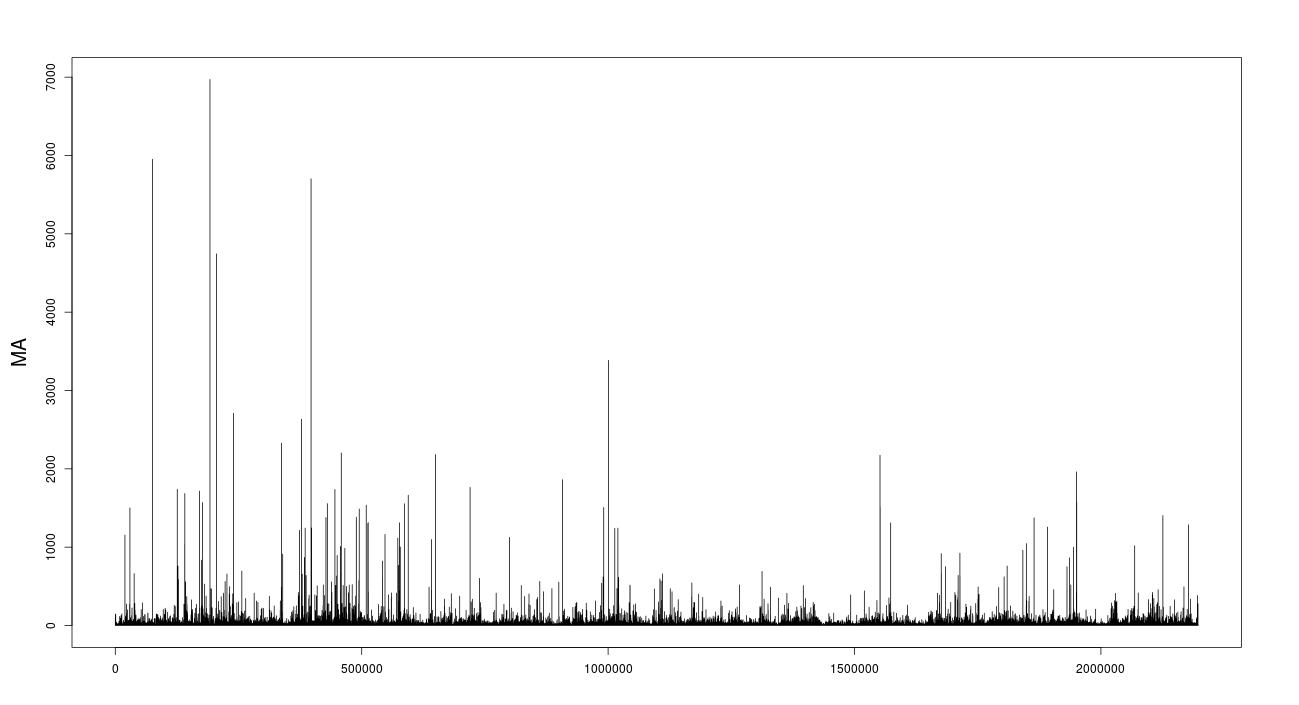}}\ \subfloat[\label{fig:6a}]{} \\[\topskip]
\raisebox{\dimexpr-.5\height-1em}{\includegraphics[width=0.99\textwidth,height=5.4cm]{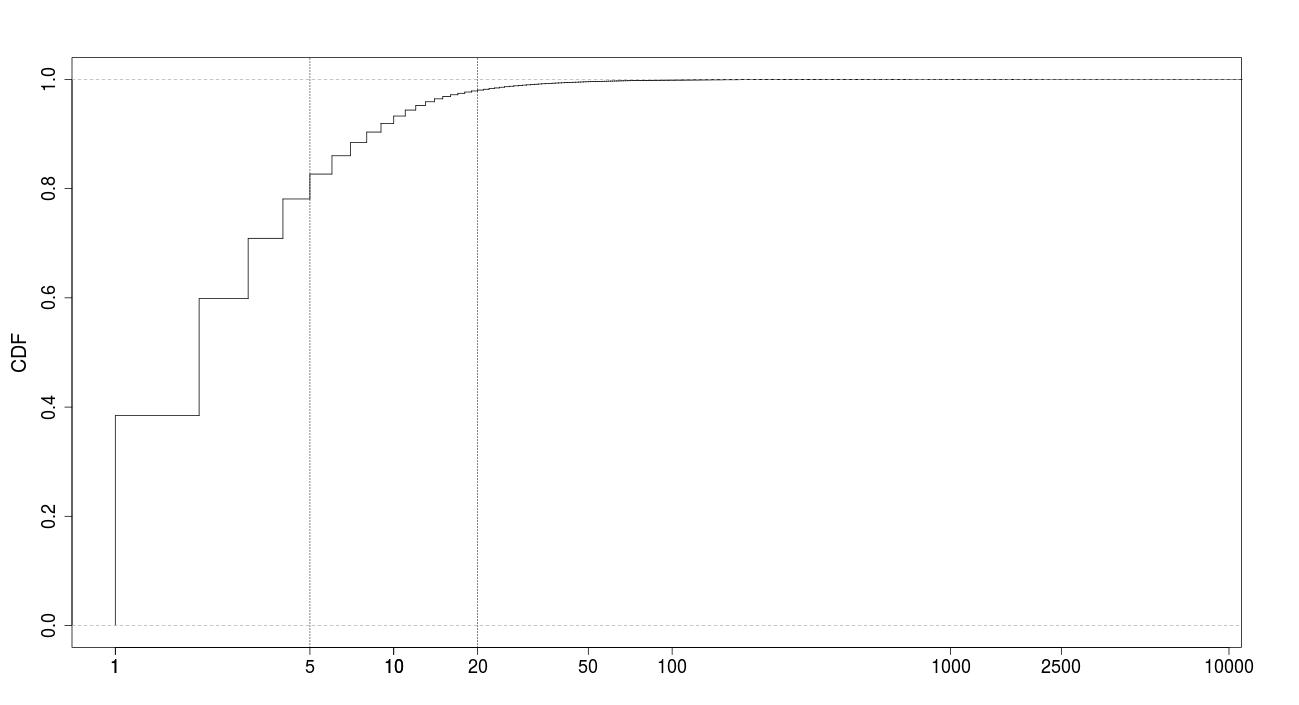}}\ \subfloat[\label{fig:6b}]{}
\caption{Number of Sessions per User (Daily) \protect\subref{fig:6a} Moving Average, and \protect\subref{fig:6b} CDF}
\label{fig:6}      
\end{figure*}

\subsection{Large Dataset}
\label{dataset}

For the current study, we use RADIUS authentication log data collected at the hotspot of the Faculty of Engineering of the University of Porto (FEUP). The University hotspots are part of the Eduroam European wireless academic network initiative. The trace data consists of the daily summary of connections between 364 APs and their corresponding wireless stations collected in almost two years, from January 1, 2010 to December 22, 2011. The university campus contains over 30 buildings, including classrooms, administrative offices, auditoriums, libraries, cafeterias, laboratories, etc. During the mentioned period, the usage record of more than 45 thousand users was observed through the established connections of over 24 million sessions. Table \ref{tab:2} depicts the evolution of the usage across the hotspot throughout the academic semesters. 

In general, an increasing trend is observed in the number of deployed APs, number of wireless users and overall number of RADIUS sessions (start, alive, and stop), from semester to semester. Total input and output traffic, however, fluctuate between spring and fall semesters to some extent. Although the overall sent and received traffic grows in volume in ultimate fall/spring semester rather than the earlier, the wireless network are subjected to higher traffic in spring semesters compared to fall semesters. 

\subsection{Preliminary Data Analysis}
In this section we present some extensive statistical analysis about the entire dataset and demonstrate relevant graphics revealing some general facts of underlying usage pattern of FEUP wireless network. We conduct this study from two peculiar viewpoint, users and the accompanying sessions, and APs and their accommodated users. 

\subsubsection{User Sessions}

As indicated earlier, each user could connect to the same AP more than once during the day, and each connection creates a separate sessionID in the accounting table. An ideal association to the wireless network could last for the entire day and if the user is fixed in its location, it is expected to have the same session without interruption. However, this is not always the case and users disassociate from their current AP and associate to the same AP or another AP in the vicinity for various reasons. Figure \ref{fig:4} considers the proportion of the user sessions in an hourly period and Figure \ref{fig:6} reveals the same information on a daily basis. 

Figure \ref{fig:4a} shows the moving average of the number of sessions that each client (device) creates during one hour of connection. Although the majority of users have a few number of sessions in an hour which shows few number of disassociations, the extreme cases are also detectable in this figure. For instance, users are observed that generate over 2000 sessions on average in an hourly connection to a single AP. To study the greatest population of users, Cumulative Distribution Function (CDF) of users and their containing sessions is demonstrated in Figure \ref{fig:4b}. This figure displays that more than 70\% of the user connections remain unbroken and preserve a single session during the hourly association to their affiliated AP, and over 95\% of the user connections contain only 5 sessions during an hour which is the result of intentional or unintentional disassociation from the current AP.

Figure \ref{fig:6a} encloses similar information as Figure \ref{fig:4a}, but in a daily basis. As expected, the number of disassociations during one day is higher than an hour period. Figure \ref{fig:6b} demonstrates that extremely consistent connections which hold a single session during a day, are less than 40\%. Most of such connections could be issued from stationary idle devices in vacant locations of the campus with few or no other active users around. This figure also displays that about 20\% of the sessions are interrupted between 5 and 20 times a day.   

\noindent
\begin{figure*}
\centering
\includegraphics[width=0.97\textwidth,height=5.8cm]{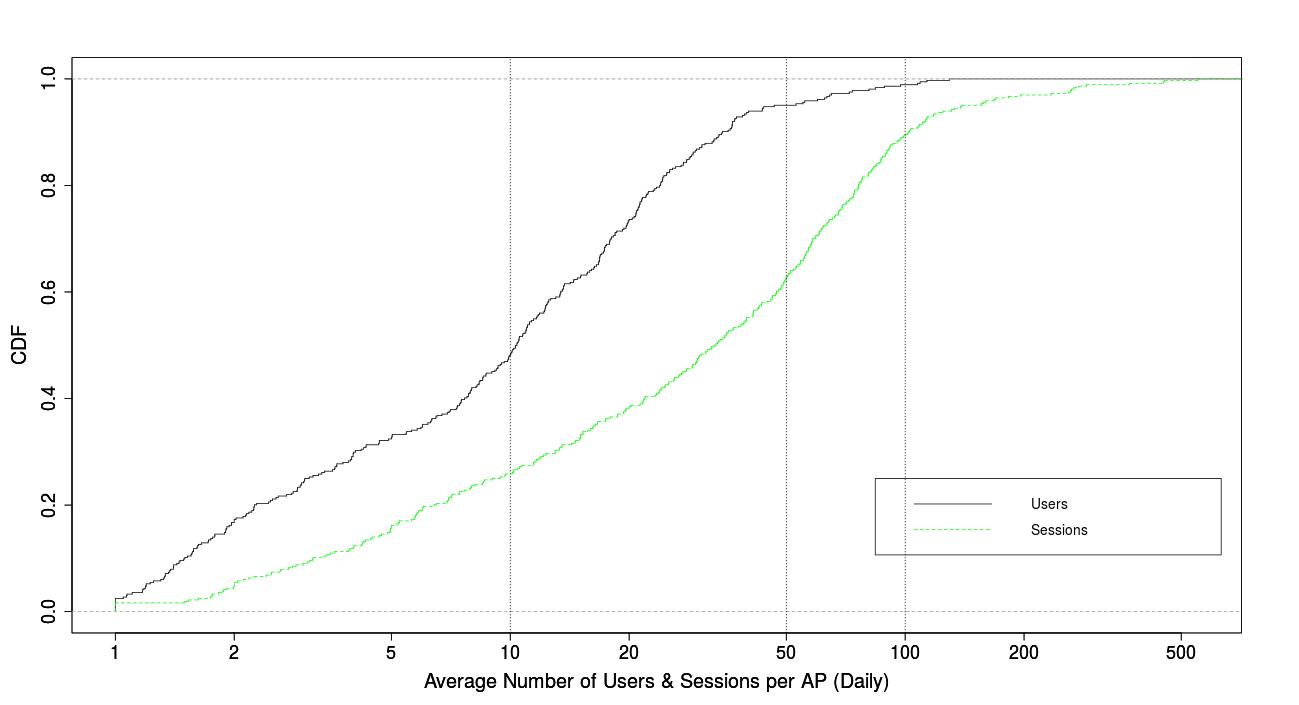}
\caption{CDF of Average Number of Users \& Sessions per AP}
\label{fig:10}      
\end{figure*}

\noindent
\begin{figure*}
\centering
\includegraphics[width=0.97\textwidth,height=5.8cm]{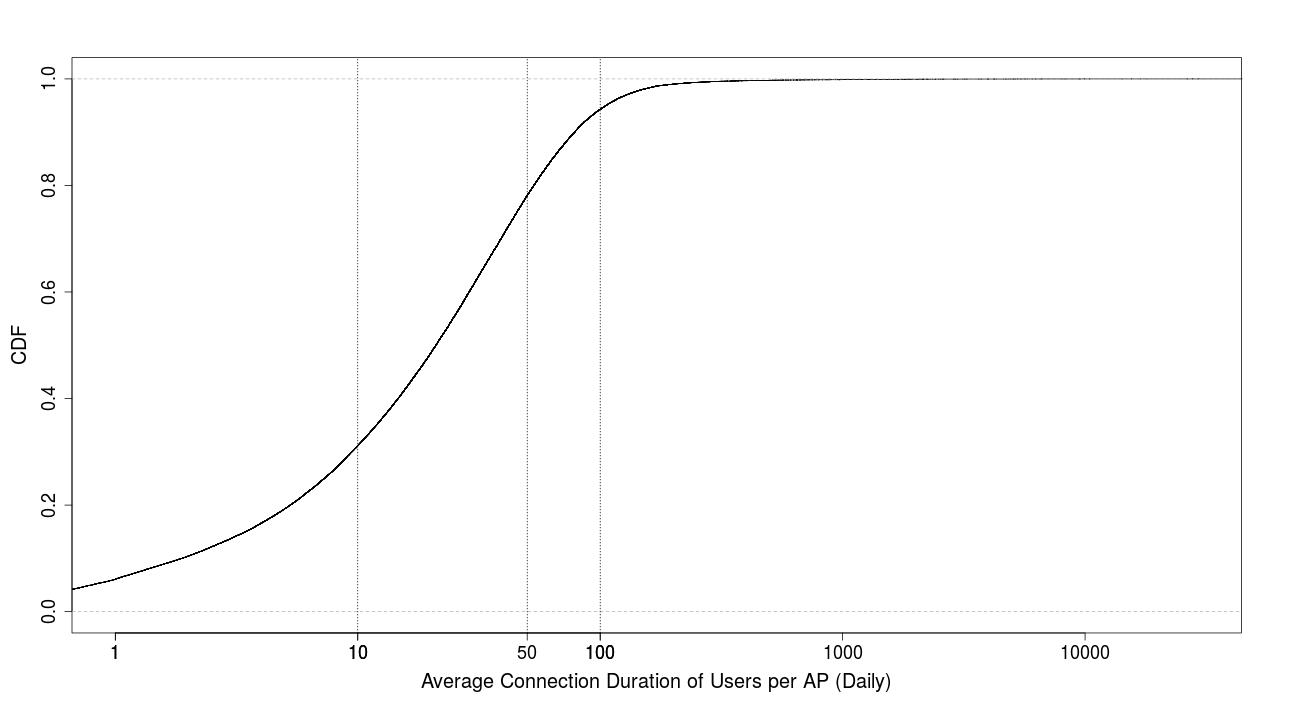}
\caption{CDF of the Daily Average Connection Duration of Users per AP (min)}
\label{fig:11}      
\end{figure*}

\subsubsection{Access Points}

In this part, the study is more focused on the usage behavior of APs as indicators of different locations around the university campus. Figure \ref{fig:10} demonstrates the average number of users and sessions per AP during the two years of experiment for the working days only. Clearly this statistics could differ from semester to semester as the number of users and their corresponding sessions evolve over time, however this figure provides a general report of involvement of the entire set of APs in the wireless covered area in two years of experiment. Figure \ref{fig:10} displays that around 20\% of the APs contain only 10 sessions per day and almost 45\% of the APs associate with 10 users during the day. To maintain the steadiness of the results, the weekends are excluded from this statistics. The figure also shows that over 95\% of the APs (345 APs) contain at most 50 users a day and about 30\% of the APs (109 APs) typically associate with 5 users each day.

Figure \ref{fig:11} reveals interesting information on the duration of users' daily connections per AP. It shows that the average connection period of users in 30\% of the time is only 10 minutes per day. This data most probably belongs to the mobile users, guests, short-term clients or inactive users. Figure \ref{fig:11} also demonstrates that around 95\% of the time, users maintain their connections to APs at most for 100 minutes, less than 2 hours a day. Such information send an important message to the network managers of the vitality of connection performance and quality of service as a great number of the users are connected to the network for less than 2 hours a day and getting interrupted over and over again in such a short period of time could be disappointing. 

The study of the information provided in Figure \ref{fig:10} and \ref{fig:11}- yields more precise understanding of the importance of the APs and learning their usage pattern based on the locations, for instance whether they are located in a busy entrance hall or a quiet corner of the campus. Such sort of information also imply the potential categories in terms of university divisions like administrative office, classroom, cafeteria, auditorium, etc. Such classification plays an important role for further analysis and modeling practices for the purpose of anomaly detection. It brings about the question of similarities (or differences) of the usage patterns in potential groups with different population of users that could prompt interesting anomaly detection strategies by learning the trend of the group and detecting the unusual events. These lines of research are of our interest for the future work.  

\subsection{Data Features}
\label{data-feat}
A number of features emerge from the raw dataset as a result of a preliminary analysis and enumeration process on a timely basis of 15 minutes. We categorize all the measured features as two main classes: \textit{Density Attributes} and \textit{Usage Attributes}. Those features that are indicators of \textit{density}, basically demonstrate how crowded is the place in terms of active attendant users, when in fact the \textit{usage} features disclose the volume of sent and received traffics by the present users. The former attributes mainly characterize the association population and durability, and the later ones reveal the total bandwidth throughput regardless of how populous is the place and it is more relevant to the applications utilized by the current mobile users. 
 
\subsubsection{Density Attributes} 

\paragraph{User Count}: the number of unique users observed in a specific location (indicated by an AP) during the predefined time-slot (15 min). 

\paragraph{Session Count} : the total population of active sessions during a time-slot regardless of the owner user. This attribute reveals the number of attempts made by the the congregation of the present users to associate to the current AP. The connection time span of each user consists of one to many sessions. 

\paragraph{Connection Duration}: the total duration of association time of all the current users. This attribute is an indicator of the overall connection persistence. The utmost amount of this features is achieved when there is no evidence of disassociation in the ongoing active sessions during a time-slot ($ User\,Count * 15\,min$).

\subsubsection{Usage Attributes} 

\paragraph{Input Data in Octets}: the number of octets transmitted from the client and incoming to the NAS port, and is only present in the Stop or Alive sessions. This attribute briefly refers to the number of bytes uploaded by the wireless user. 

\paragraph{Output Data in Octets}: the number of octets received by the client and leaving the NAS port, and is only present in the Stop or Alive sessions. This attribute shortly refers to the number of bytes downloaded by the wireless user. 

\paragraph{Input Data in Packets}: the number of packets transmitted by the client and incoming to the NAS port. This attribute is similar to the above \textit{Input-Octet}, just to be measured in packets instead of bytes.

\paragraph{Output Data in packets}: the number of packets received from the client and leaving the NAS port. This attribute is similar to the above \textit{Output-Octet}, just to be measured in packets instead of bytes.


\noindent
\begin{figure}
\centering
\includegraphics[width=0.5\textwidth]{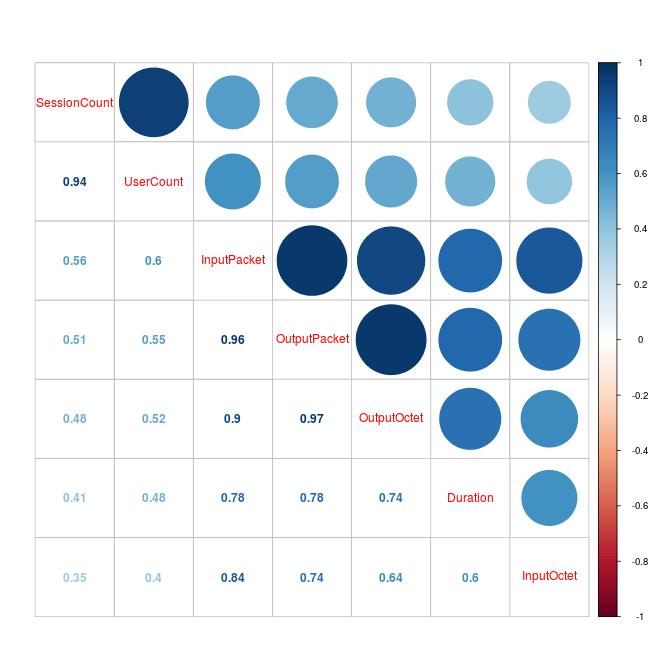}
\caption{Correlation Matrix of the Main Data Features}
\label{fig:1}      
\end{figure}

\noindent
\begin{figure*}
\centering
\includegraphics[width=0.8\textwidth]{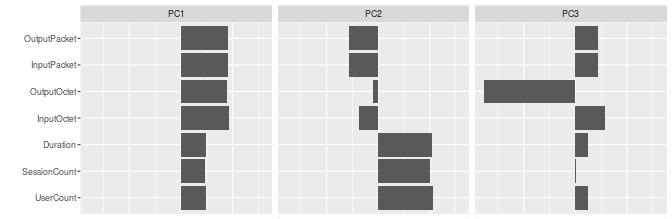}
\caption{The Behavior of the Main Features Relative to the Three Principal Components}
\label{fig:3}      
\end{figure*}

\subsection{Feature Selection}
\label{feat-sel}

In this section we discuss the connection and correlation of the data features explained earlier and disclose how to choose the best set of features for further analysis. 

Figure \ref{fig:1} depicts the correlation matrix of all the above features. There is a high correlation observed between \textit{User Count} and \textit{Session Count}, on the grounds that the number of sessions are always equal or higher than the number of users in a time-slot. \textit{Duration} do not have a strong correlation with any of the mentioned features, neither with \textit{Density Attributes}, nor with \textit{Usage Attributes}. 

Having considered the input and output traffic transferred in octets, there is no significant correlation between these two compared to Input and output data in packets. However there is a slightly noticeable correlation between \textit{Output Octets} and its corresponding attribute \textit{Output Packets}, as well as \textit{Input Octets} and \textit{Input Packets}. Although there is a slight correlation between input/output data in octets and in packets, we consider them as semi-independent variables and include both of them in our further experiments. The information added to the system through input and output traffic in octets simply take into account all the sent and received data in bytes. However, the input and output traffic measured in packets, could bring other types of information as the packets' size could differ by various factors such as application types and communication protocols.

For subsequent analysis and modeling procedures, we favor using less features rather than the entire set of attributes introduced earlier. For this reason, we applied Principal Component Analysis (PCA) technique to find the combination of the variables which best explain the phenomena and contain the greatest part of the entire information. 

In this case the first three principal components bring the cumulative proportion of variance to over 95\%. Figure \ref{fig:3} demonstrates the participation proportion of each feature to the principal components.
From an analysis of this Figure we conclude that the first principal component is associated with all the above features in a positive manner, more specifically with the usage attributes. The second principal component is declined towards the density attributes, increasing with the larger density values and yet decreasing with the larger usage values. The single largest contributor to the third principal component is the input data in octet or the amount of uploaded bytes by the wireless users. The other features play less important roles in the third component, positively or negatively. Approximately categorizing the principal components like so, provides us with a deeper understanding of the connection of the aforementioned features, density or usage attributes, with the emanate best features resulted by PCA technique.

\subsection{Conclusions}
In this section we introduced collected RADIUS data from FEUP hotspot as the main dataset of this work and performed a preliminary analysis from two points of view- user sessions and access point- to demonstrate the situation of the data respecting the hourly and daily sessions per user as well as the user population and connection duration per AP. Moreover we presented two main groups of features- usage attributes and density attributes- and defined a number of features for each group. We further studied the connection and correlation of the data features and selected the best combination of those features applying PCA. In the upcoming section we show how to model the AP usage data using the selected features represented in this section. 

\section{Statistical Modeling of 802.11 AP Usage}
\label{modeling}
In this section we introduce statistical techniques for modeling purposes and in the upcoming section we indicate how to apply these models for anomaly detection. The modeling approach itself can be used in distinct directions such as to study the similarities and differences of the locations, to categorize the localities in terms of functionality (e.g. classroom, office, library) or specification (homogeneous/heterogeneous daily, seasonal or constant usage). We introduce time-invariant and time-variant models and in each case we show how to apply the model on the large dataset previously elaborated.  

\subsection{Time-invariant Modeling}
\label{time-invar}
We first consider models that assume there is no time binding between consecutive daily events. Although this might not be precisely the case, it yields simpler modeling approach. Later in the paper we compare this type of modeling with others that do consider dependency between consecutive daily events.

\subsubsection{Gaussian Mixture Model}
\label{gmm}
We begin our modeling efforts by applying techniques that assume all daily events come from the same distribution, regardless of any time dependency between the consecutive records. To explain this, we pick Gaussian Mixture Model (GMM), a probabilistic model that presume all the data points are generated from a mixture of a finite number of Gaussian distribution with unknown parameters. The Expectation Maximization (EM) procedure is the optimization technique utilized to fit the unknown parameters and incorporate information about the covariance structure of the data as well as the centers of the latent Gaussians \cite{Ref31}.   

\begin{equation}
  p(x|\lambda) = \sum _{k=1}^{M} \omega _{k} \, g(x|\mu _{k},\Sigma _{k})
\label{equ:gmm-1}
\end{equation}

where $ x $ is a D-dimensional continuous-valued data vector (of features), $ w_{k}, k = 1, . . . , M $, are the mixture weights, and $ g(x|\mu _{k},\Sigma _{k}), k = 1, . . . , M $, are the component Gaussian densities. Each component density is a D-variate Gaussian function of the following form,

\begin{equation}
  g(x|\mu _{k},\Sigma _{k} ) = \frac{exp \{-\frac{1}{2} (x-\mu _{k})'\Sigma _{k}^{-1} (x-\mu _{k})\}}{(2\pi )^{D/2} |\Sigma _{k}|^{1/2} }  \quad
\label{equ:gmm-2}
\end{equation}

with mean vector $ \mu _{k} $ and covariance matrix $ \Sigma _{k} $ .The mixture weights satisfy the constraint that $ \sum _{k=1}^{M} \omega _{k} = 1 $.

The complete Gaussian mixture model is parameterized by the mean vectors, covariance matrices and mixture weights
from all component densities. These parameters are collectively represented by the following notation,

\begin{equation}
  \lambda = \{ \omega _{k},\mu _{k},\Sigma _{k}\}  \qquad   k = 1, . . . , M
\label{equ:gmm-3}
\end{equation}

\subsubsection{GMM Application: Case Study}
\label{gmm-case-study}

GMM could be applied to our data features in several ways, for instance a single mixture model for the entire set of data, or a mixture model for each location separately. The later approach is closer to our goal of proposing practical models for each place indicated by an AP (or a broader neighborhood) to explore the characteristics of that place, and ultimately discovering the abnormal behaviors occurring in contrast with the expected usage pattern. Note that in our previous work \cite{Ref24} we modeled and identified the anomalies of three categories: of a single model for all APs, a mixture model for groups of APs and individual models for each AP. In this work we study the individual model to be able to evaluate it with our deployed testbed and in the future work we intend to explore the models for the potential groups of APs.   

In order to investigate the modeling capacities of GMM for the mentioned aims, we select two different spots to be our test cases: a highly crowded AP at the computer service section with 3726 observed users, and a less crowded AP in the chemical engineering department with overall 175 users. The experiment takes into consideration the second semester period of 2011 from February to July. To achieve more precise result, we focus on the working daily pattern, hence the data records belong to the working days (from Monday to Friday) and the working hours (8 a.m. to 6 p.m.).

\begin{figure*}
\centering
\includegraphics[width=1.0\textwidth,height=11cm]{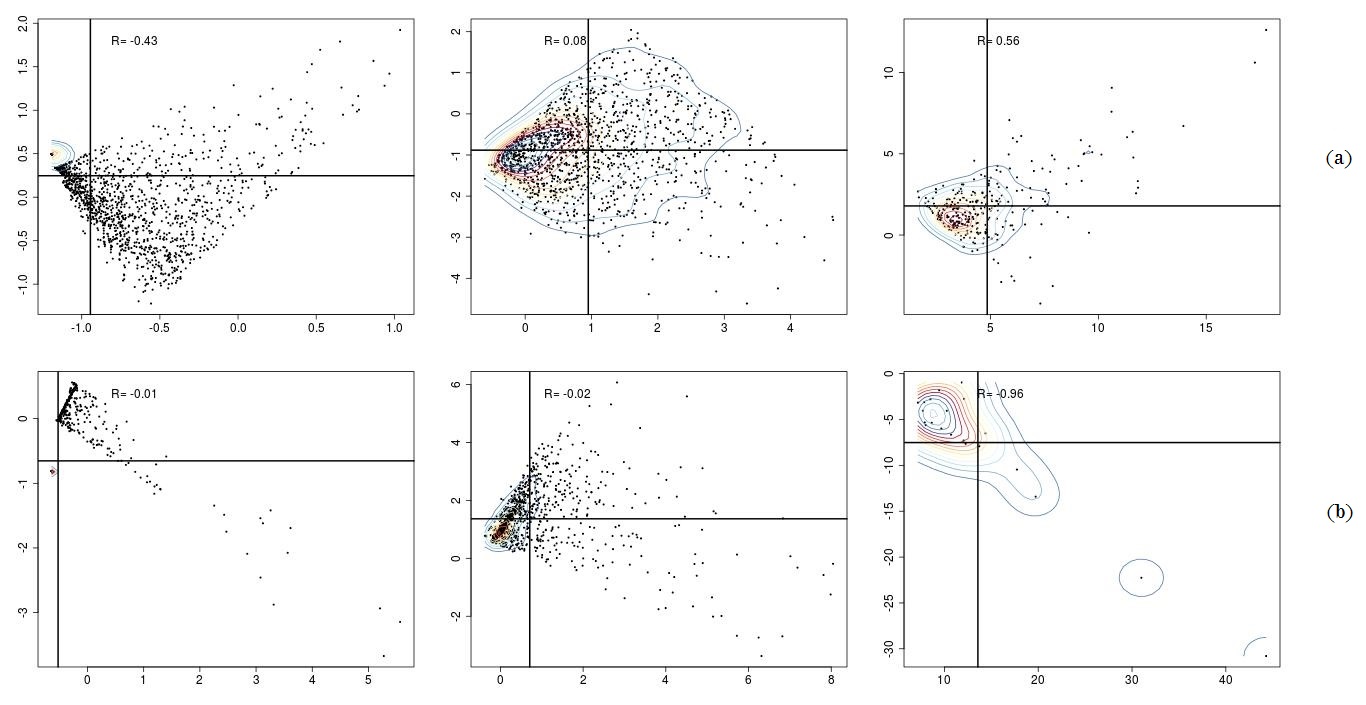}
\caption{Density Parameters of Three Gaussian Mixture Components of the Selected APs. \textbf{a} Crowded AP, \textbf{b} Less Crowded AP}
\label{fig:20}      
\end{figure*}

On each location, GMM fits are computed with three mixture components. The Gaussian density parameters (mean and covariance matrix) are depicted in Figure \ref{fig:20}, the first row belongs to the crowded AP and the second row shows the density parameters of the less crowded AP. In order to facilitate the visual perception and to have an easier comparison, the density parameters are illustrated in 2D, despite the fact that GMM process is conducted on 3 features (principal components). 
 
The data is standardized on each column to have zero mean and one standard deviation, so the density values are not appropriate to be compared with each other directly. However, the contour lines show the diversity of the data points in each mixture component and the direction of spread as well as the mass center. The R value on each plot represents the correlation between the X and Y axis, correspondingly the first two principal components. 

Each location is characterized in this manner and according to GMM modeling approach,

$ \lambda_{1} = \{ \omega _{i1},\mu _{i1},\Sigma _{i1}\} \qquad i = 1, . . . , 3 $ 

and 
 
$\lambda_{2} = \{ \omega _{j2},\mu _{j2},\Sigma _{j2}\} \qquad j = 1, . . . , 3 $ 

represent the mixture weights and density parameters of the first and the second APs respectively. 
 
\noindent


\subsection{Time-variant Modeling}
\label{time-var}
In this section we consider models that assume time dependency between consecutive daily events. In this case the sequences of data records matter and they form significant connections in a meaningful context or profile. In time-variant models in general, conditional probabilities for events are determined based on the history of the events. In the following section we study the Hidden Markov Models for modeling the time-varying sequential data for the ultimate purpose of anomalous pattern recognition which we discuss more in detail in the next section.         

\subsubsection{Hidden Markov Model}
\label{hmm}
HMMs are generally used for the stochastic modeling of non-stationary time-series. HMMs provide a high level of flexibility for modeling and analyzing time-varying processes or sequential data. Their particular application is in recognition such as speech recognition, activity recognition, gene prediction, etc. where data instances are represented as a timely sequence of estimates. In the current research we propose how to use HMMs for modeling and anomaly detection purposes in wireless networks which has never been investigated before to the best of our knowledge.

Rabiner and Juang \cite{Ref30} presented a comprehensive tutorial on HMM which provides a profound understanding of the basic blocks of HMM. HMM symbolizes a doubly stochastic process with a set of observable states and a series of hidden states which can only be observed through the observable set of stochastic process. The goal in HMM is recovering a data sequence that is not immediately observable through the other set of observable data.

\noindent
\begin{figure*}
\centering
\includegraphics[width=1.0\textwidth,height=11cm]{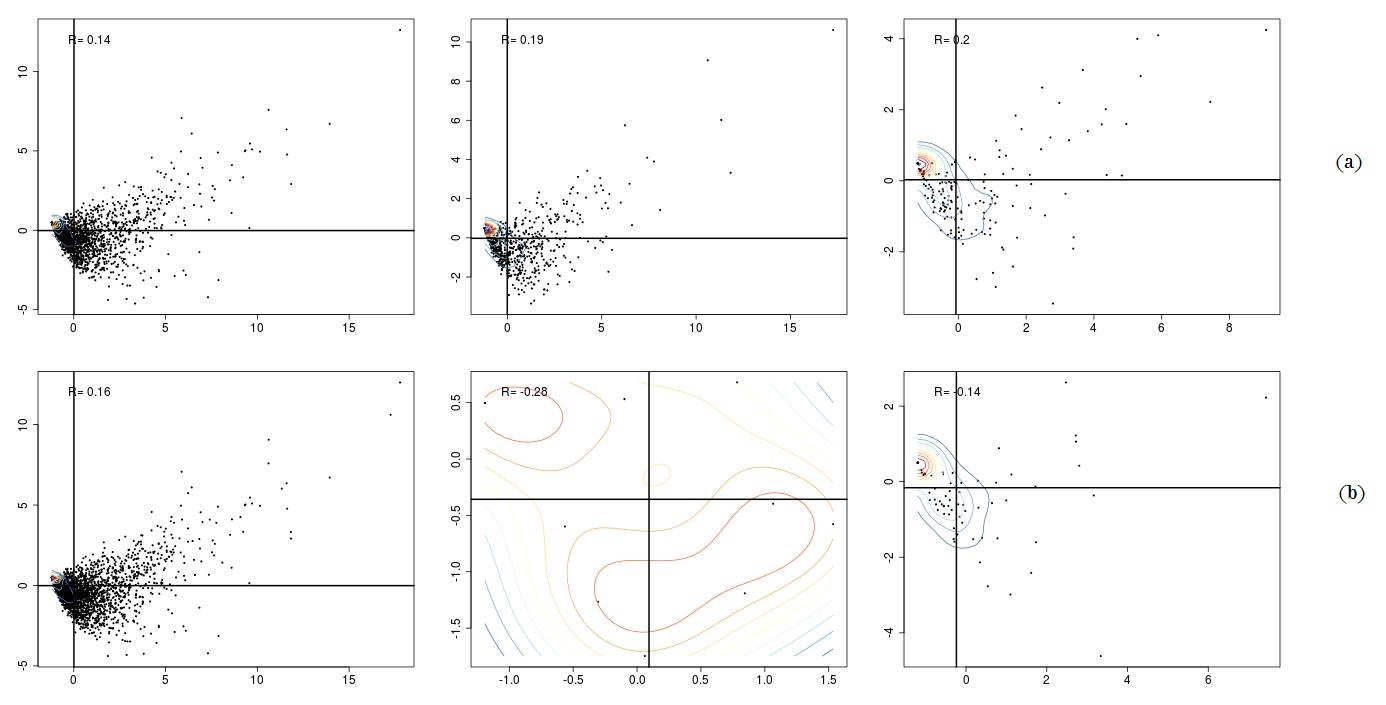}
\caption{Density Parameters of Three Hidden State in HMM of the Selected APs. \textbf{a} Crowded AP, \textbf{b} Less Crowded AP}
\label{fig:21}      
\end{figure*}

The formal definition of a n-state HMM notation is determined as follows:
 
\noindent
\begin{itemize}
  \item A set of hidden states $S = \{s_{i}\}$	,	$1 \le i \le n$
  \item A set of possible symbol observations in discrete models $V = \{v_{i}\}$	,	$1 \le i \le m$
  \item State transition probability distribution (transition matrix) $A = \{a_{i,j}\}$	, $1 \le i,j \le n$ , $a_{i,j} = P(s_{j}$ at $t+1 | s_{i}$ at $t)$
  \item Observation symbol probability distributions (emission matrix), $B = \{b_{j}(k)\}$ ,	$1 \le k \le m $ , $b_{j}(k) = P(v_{k}$ at $t | s_{j}$ at $t)$ 
  \item Initial state distribution $ \pi = \{\pi_{i}\} $,	$1 \le i \le n$ , $\pi_{i} = P(s_{i}$ at $t=1)$
	\item $m = $ number of observation symbols in discrete models
	\item $n = $ number of hidden states
\end{itemize}

The set $ \lambda = (A, B, \pi) $ completely defines an HMM \cite{Ref30}. However, in continuous emissions, instead of having $m$ outcomes for the observations, distribution parameters such as mean and covariance are determined. In such cases a model is represented as $ \lambda = (A, \mu ,\Sigma , \pi) $, and $\mu$ and $\Sigma$ stand for mean vector and covariance matrix respectively. 

Using the model $ \lambda$, an observation sequence $ O = o_{1},o_{2},...,o_{T} $ is generated as follows:

\noindent
\begin{enumerate}
  \item Select an initial state, $s_{1}$, according to the initial state probability distribution, $ \pi $;
  \item Set $ t=1 $;
  \item Choose $ o_{t} $ according to observation probability distribution in state $s_{t}$, $b_{s_{t}}(k)$;
  \item Choose $ s_{t+1} $ according to the state transition probability distribution for state $s_{t}$, $ a_{s_{t},s_{t+1}} $
  \item Set $ t=t+1 $; return to step 3 and continue until $ t>T $
\end{enumerate}

Given the form of the HMM, there are three key problems of interest that solving them promotes modeling the real world applications. These problems are listed as the following \cite{Ref30}:

Problem 1 -- Given the observation sequence $ O = o_{1},o_{2},...,o_{T} $ and the model $ \lambda = (A, B, \pi) $, how we compute $ P(O|\lambda) $, the probability of the observation sequence.

Problem 2 -- Given the observation sequence $ O = o_{1},o_{2},...,o_{T} $, how we choose a state sequence $ S = s_{1},s_{2},\\...,s_{T} $ which is optimal in some meaningful sense.

Problem 3 -- How we adjust the model parameters $ \lambda = (A, B, \pi) $ to maximize $ P(O|\lambda) $. 

According to our data set, the HMMs form observations with continuous multivariate Gaussian distribution, hence the emission matrix $B$ is defined by the distribution parameters associated with the set of states. In the proposed model, the HMMs contain fully connected states, thus transitions are allowed from any state to any other state. 

\subsubsection{HMM Application: Case Study}
\label{hmm-case-study}

In this section we select the very same APs as in the GMM case study (Section \ref{gmm-case-study}), and build HMM models for each of them separately. Our focus is once more on the working daily pattern in the second semester of 2011, from Monday to Friday in the working hours.

As described earlier, we consider fully connected HMMs (ergodic model) with continuous Gaussian distribution as the emission probabilities and 3 hidden states. The states are initialized randomly, and the number of states is selected heuristically based on the best practice of the experiments conducted on both the large dataset and the Testbed dataset. For the multivariate Gaussian distribution of the observations, each component of the mean vector is uniformly drawn between $\mu - 3\sigma $ and $\mu + 3\sigma $ and the initial covariance matrix is diagonal and each initial variance is uniformly drawn between $\frac{1}{2}\sigma ^{2}$ and $3\sigma ^{2}$. The initial probability matrix ($\pi$) and the transition matrix ($A$) are 
 uniformly drawn. The initial HMM is then optimized with the Baum-Welch algorithm with the cut off likelihood value of $1e-6$ or the maximum number of iterations set to 20. After the optimization process, the physical meaning of the hidden states are more discernible. The values of the principal components in each state shows the tendency of the states to the usage or density attributes. For example a hidden state with the highest value for the second principal component shows a more populated case in terms of users or sessions density. In the future work where the concern is more on modeling the anomalous patterns we utilize the interpretation of the hidden states to relate them to the physical conditions of the locations.   

The Gaussian density parameters of the three hidden stated are illustrated in Figure \ref{fig:21}, similar to Figure \ref{fig:20}, the first row is affiliated with the crowded AP and the second row belongs to the less crowded AP. The contour lines in these two figures represent the overall picture of the population and density distribution of the data in each component or state. Suchlike graphs are visual aids to depict the density parameters only, and for inspecting the goodness of distribution over the entire feature set and make any  comparison, further investigations are required. 


\subsection{Model Comparison: GMM vs. HMM}
In this section two techniques are considered only for the sake of modeling purposes, a time-invariant model (GMM) and a time-variant model (HMM). In the coming section we investigate the ultimate goal of this modeling which is the recognition of anomalous points or regions. At this stage, before exploring the anomaly detection territory, we briefly itemize the modeling functionalities and propose some simple tests to verify the more qualified model.

The potential functionalities of the locations characterization and modeling are listed as following:
\begin{itemize}
   \item Classification of the locations, represented by APs, in terms of utility and temporal patterns.
   \item Recognition of the meaningful similarities and distinction of the locations.
   \item Grouping the most related APs and propose mixture models for the groups \cite{Ref24}.
\end{itemize} 

%

\begin{table*}[!htbp]
\centering
\caption{Log-likelihood Values (LLVs) of the Training and Test Data Belong to the Selected APs for GMM and HMM Models}
\label{tab:3}
\begin{tabular}{|l|c|c|c|c|}\hline
\diaghead(-5,1){\hskip5.2cm}%
    {Test Data LLVs}{Trained Model}&
\thead{GMM \\ Crowded AP}&\thead{GMM \\ Less Crowded AP} & \thead{HMM \\ Crowded AP}&\thead{HMM \\ Less Crowded AP}\\
\hline \hline
The same train data & -3468 & -2154 & -2553 & -2131 \\ \hline
Test data from the crowded AP & -189 & -189 & -134 & -209\\ \hline
Test data from the less crowded AP & -509 & -95 & -195 & -115\\ \hline 
\end{tabular}\medskip
\end{table*}
 
To investigate the competency of the two proposed models and estimate the capacity of each, we conduct a simple test. First of all, we measure the log-likelihood of the models in modeling the training data of the two samples, crowded AP and less crowded AP, and then we select a random day from each AP and calculate the log-likelihood of the models towards the test data which is new to both models. 
We use log-likelihood values (LLV) to measure the goodness of fit of our models. The model with larger log-likelihood value surpasses the model with smaller log-likelihood value. 

Given data $x$ with independent multivariate observations $x_{1},...,x_{n}$, the likelihood of a Gaussian mixture model with $M$ components is defined as \cite{Ref32}:

\noindent
\begin{equation}
likelihood(x|\lambda) = \displaystyle\prod _{i=1}^{n} \displaystyle\sum _{k=1}^{M} \omega _{k} \, g(x_{i}|\mu _{k},\Sigma _{k}) 
\label{equ:gmm-ll}
\end{equation} 

where $g(x|\mu _{k},\Sigma _{k})$ is the $k$th component's Gaussian density, as already defined in Equation \ref{equ:gmm-1}, and $\omega _{k}$ is the probability that an observation belongs to the $k$th component. 

The log-likelihood function takes the following form:

\noindent
\begin{equation}
$log-likelihood$(x|\lambda) = \displaystyle\sum _{i=1}^{n} log(\displaystyle\sum _{k=1}^{M} \omega _{k} \, g(x_{i}|\mu _{k},\Sigma _{k})) 
\label{equ:gmm-logll}
\end{equation} 

In the EM process, the parameters of the GMM, $\lambda$, are estimated so that the likelihood of the GMM given the training data is maximized, Maximum Likelihood Estimation (MLE). Ensuing several iterations, the MLE yields the likelihood of the GMM given the training data. We applied MClust R package \cite{Ref33} to conform the Gaussian mixture components and estimate the log-likelihood of the training and test data provided in Table \ref{tab:3}.   

The likelihood of a HMM is basically the first key problem of HMMs stated earlier, the probability of an observation sequence given the model parameters:

\noindent
\begin{equation}
\begin{array}{r c l}
P(O|\lambda) &=& \displaystyle\sum _{all \, S} P(O|S,\lambda) P(S|\lambda) \\
             &=& \displaystyle\sum _{s_{\!1\!},s_{\!2\!},...s_{\!T\!}}\pi_{s_{\!1\!}}b_{s_{\!1\!}}\!(\!O_{1}\!)a_{s_{\!1\!},s_{\!2\!}}b_{s_{2}}\!(\!O_{2}\!)...a_{s_{\!T\!-\!1\!},s_{\!T\!}}b_{s_{\!T\!}}\!(\!O_{T}\!) 
\end{array}
\label{equ:hmm-ll}
\end{equation} 

We utilized GHMM library \cite{Ref34} for the formation of HMMs, estimation of log-likelihoods and all the other requisites of the experiments performed in this work. 

Table \ref{tab:3} contains the log-likelihood values of the trained GMM and HMM models for the selected APs, regarding both the training and test data. Comparing the log-likelihood values of the training data, HMM provides higher values (less negative) both for the crowded AP and the less crowded AP. Note that the training data contains 25 days data and the test data consists of only one day data selected randomly from the unobserved days. Concerning the test data, it is expected that the selected day from the same AP obtains higher log-likelihood value rather than the data from another AP due to the possible similarity of daily usage in a specified location. The first GMM (built over the crowded AP data) provides the same amount of log-likelihood for both test data, thus yields no distinction for its own usage pattern rather than the other AP. However, the second GMM (trained with the less crowded AP data) provides higher log-likelihood value for its own data rather than the other AP. 

HMMs, on the other hand, provide higher amount of log-likelihood for their own test data rather than the other AP, which shows the better matched model for self data. Both GMM and HMM models for the crowded AP provide close values of log-likelihood for the test data, so the models do not seem to be very robust in distinguishing between its own data and the other AP. Howbeit, GMM and HMM models for the less crowded AP achieve higher log-likelihood values for the training data rather than the models of the crowded AP. It must be considered that the test data is selected randomly and the pattern of the selected day is not determined in terms of normal or abnormal usage, nevertheless the overall outcome of HMM models looks more satisfying compared with GMM. In Section \ref{exp-setup}, the experiments are conducted on the testbed dataset with recognized anomalies so that the conclusion will be based on the known ground truth. In the next section, we investigate the time-variant specifications of HMMs towards the simplicity of the time-independent GMM concerning the anomaly detection objectives.  

\subsection{Conclusions}
In this section we presented GMM as time-invariant and HMM as time-variant modeling techniques. As a case study for each approach we selected two different locations in the university campus- a highly crowded AP and a less crowded AP- and applied the forenamed methodologies. We then defined the log-likelihood for each method separately to examine the goodness of fit for the proposed models in terms of train and test data. Having conducted a simple experiment on the selected APs revealed that HMMs are more likely to provide a robust model to distinguish between their own pattern and an unfamiliar pattern. In the following section we show the functionality of the proposed models to detect anomalous cases in AP usage data.

\section{Detection of Anomalies in AP Usage Data}
\label{anom-detect}
In this section we show how the aforementioned models are utilized for the purpose of anomaly detection. We further explore the capabilities of these models in recognition of abnormal events and series of unexpected occurrences. 

\subsection{Anomaly Detection Approach}
\label{anom-detect-appr}

Network administrators are generally concerned with anomaly detection as well as prediction. These two important tasks enable them not only to make immediate decisions to alleviate the complications of the network, but also to establish longstanding plans to support the expansion of the network and its dynamic usage over time.
 
\subsubsection{GMM Estimation: Divergence from the Gaussian Densities}
The most generic definition of the anomalies asserts those points or small regions isolated from the normal zones which contain the majority of the observations. Thus, a straightforward approach to detect anomalies, when there is no ground truth available, is to define the normal zones and distinguish those rare observations which hardly belong to those normal sectors.

In GMM, the time-invariant model discussed earlier, a number of Gaussian mixture components are determined and each component contains normal density parameters. The model is built based on several training data and the newly arrived records are inclined to the most compatible component with the least distance. Hence, to detect abnormal points we need to estimate the affinity degree of each point, as already described in Equation \ref{equ:gmm-ll}, and mark outliers as having the slightest probability of belonging to any cluster.

\subsubsection{HMM Estimation: Likelihood Series}
HMM, as a time-variant model, considers the temporal dependency between consecutive data records. Calculating the log-likelihood of a single data point or a series of sequential data points as already expressed in Equation \ref{equ:hmm-ll}, emanates the mis-behaving records comparing to the log-likelihoods of the norm of the data. The unexpected low values for the log-likelihood in HMM are generally due to one or some of the following arguments:

\paragraph{Divergence from the Assigned Hidden State:}
Given an HMM model $\lambda$ and an observation sequence of $ O = o_{1},o_{2},...,o_{T} $, the most probable set of states are generated by \textit{Viterbi} algorithm as $ S = s_{1},s_{2},...,s_{T} $, $s_{i} \in S$. To estimate the distance of a data point in time $t$ to its counterpart HMM state ($s_{t}$) in Viterbi path, the \textit{Mahalanobis} distance is evaluated between time-series elements and the hidden states. Consequently the outliers which display the unreasonable distance to their assigned hidden states, are highlighted to potentially have a poor value in the likelihood series. This approach is approximately similar to the outlier detection technique addressed earlier for GMM components.

\paragraph{Less Likely State Transition:}
According to the third well-known HMM problem, stated in Section \ref{hmm}, when a HMM model is trained the model parameters are adjusted to maximize the probability of the observed data $P(O|\lambda)$. The transition probability matrix is one of the salient components of the trained model. The highest transition probabilities are frequently observed between identical states ($s_{i}$ to $s_{i}$), while the lowest probabilities often occur between the most distant states. However, regardless of the form of the transition matrix, in the Viterbi sequence outcome, it is expected to observe the transition probabilities proportional to the values of the transition matrix. Whenever this principal is violated there exist an anomaly prospect. For instance if in a Viterbi path the transition from state $s_{i}$ to state $s_{j}$ occurs only once (out of 60 transitions), and the transition probability of $a_{i,j}$ is 30\% in the transition matrix, this circumstance sounds unlikely and thus an anomaly-prone transition.

\subsubsection{Anomaly Detection: Case Study}

\begin{figure}
\centering
\includegraphics[width=0.5\textwidth]{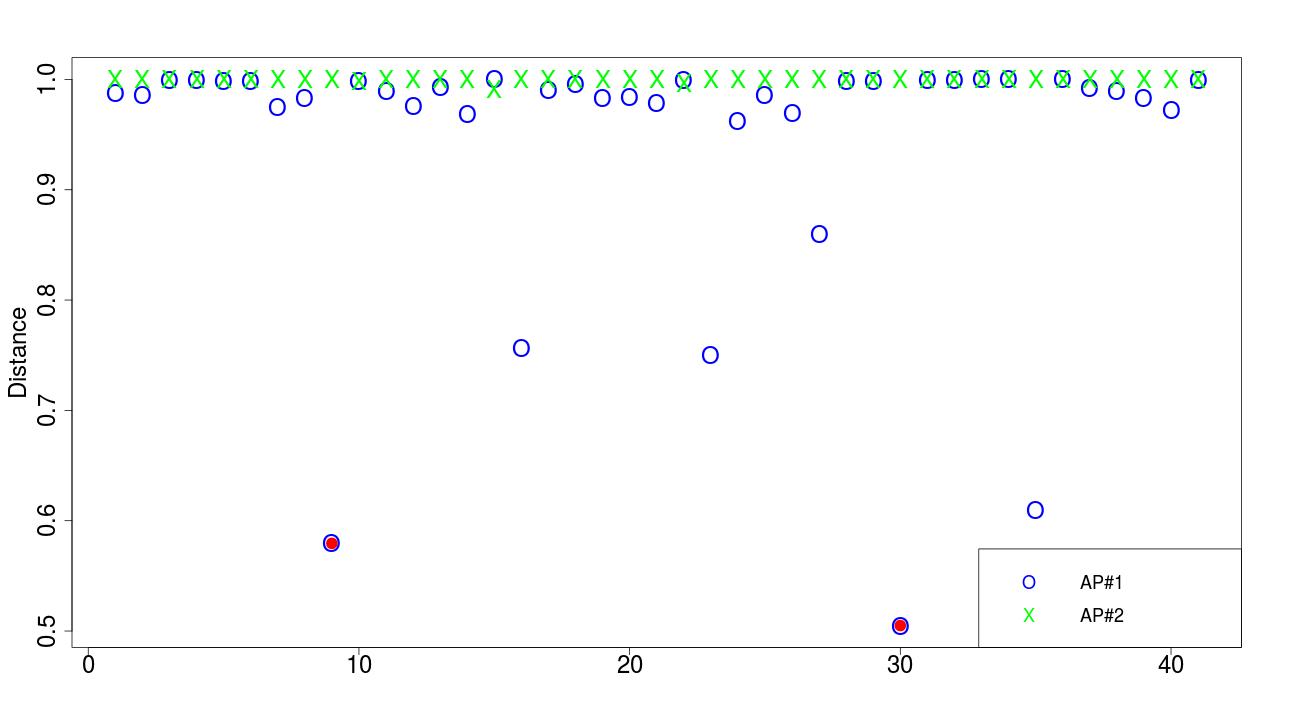}
\caption{GMM Estimation of Anomalous Data Points Based on the Largest Distance from the Assigned Gaussian Component}
\label{fig:30}      
\end{figure}

In this section we explore the addressed methodologies to detect anomalous data points or data sequences in the same two APs that we proposed GMM and HMM models for their corresponding training data. Figure \ref{fig:30} highlights the outliers of the one day test data detected by measuring the largest distance from the Gaussian components. The result of the first AP (crowded AP) is displayed in blue and the second AP (less crowded AP) is demonstrated in green. Two data points are marked in red that both belong to the first model of the crowded AP. These outliers are appointed to a Gaussian component of the first model, but with the lowest probability (less than 60\%). Here we selected the normality threshold to be 60\%, however it could differ from model to model and the most appropriate value of threshold could eventually be decided by the network manager.  

\begin{figure}
\centering
\includegraphics[width=0.5\textwidth]{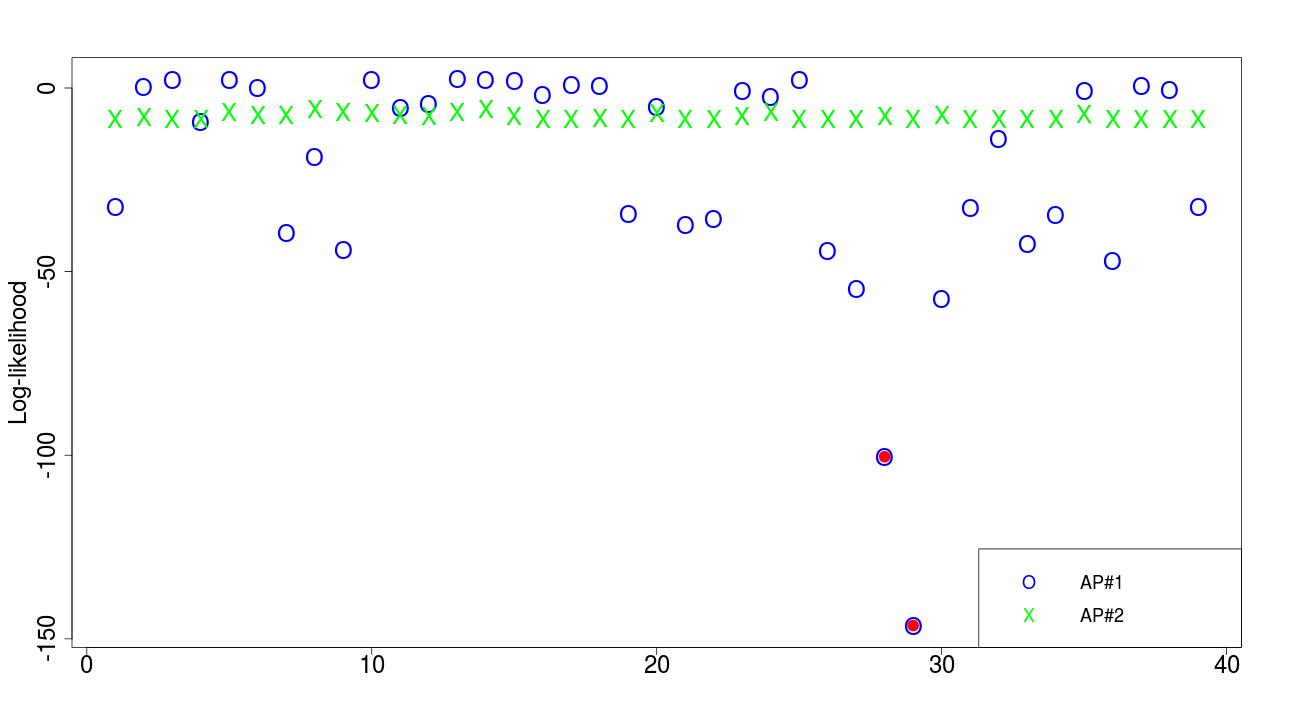}
\caption{HMM Estimation of Anomalous Data Points Based on the Lowest Log-likelihood}
\label{fig:31}      
\end{figure}

Figure \ref{fig:31} displays the anomalous points detected by HMM based on the lowest value of the log-likelihood. In this approach, two different data points are marked as outliers which belong to the first AP training data, the crowded AP. The cut-off value is considered to be log-likelihoods below -100, note that this value could also be configured. The more strict cut-off value yields higher false positive rate. We investigated the likely origins of the outliers emerged in this case and we observed that the \textit{Mahalanobis} distance of the marked data points are maximal with the assigned hidden state in the \textit{Viterbi} path. That must have caused the low log-likelihood value in the likelihood series. Further experiments on anomaly detection by HMMs and evaluation techniques are performed in our previous work in \cite{Ref24}.

\noindent

\noindent

However, in this case study we demonstrated how the anomaly detection analysis work in our proposed framework. In the next section, we evaluate both models based on the achieved results of the deployed testbed, hence we can determine with more confidence which points are detected correctly. 

\subsection{Conclusions}
In this section we described the anomaly detection techniques by GMM and HMM. In GMM we define anomalies as the distant data points that hardly belong to any Gaussian component, while in HMM anomalies are the data points with the minimum likelihood value. As discussed more in detail in our previous work \cite{Ref24}, we analyzed the root cause of the low likelihood value as divergence from the assigned hidden states as well as the low probability in state transition. We further explored the addressed methodologies to detect anomalies at the same APs from the previous section. We justified the detected anomalous points, however in absence of the ground truth in the large dataset it was not possible to throughly evaluate the anomalous points and we left the evaluation process for the next section.

\section{Experimental Setup}
\label{exp-setup}

In order to validate anomaly detection techniques proposed in this work we deployed an exploratory testbed with one single AP and generate a number of anomalies in a controlled environment for experimental purposes.
We work with FreeRADIUS server which is widely used for Enterprise Wi-Fi and IEEE 802.1X network security and communication, particularly in the academic community, including Eduroam \cite{Ref28}. The very basic aspects of our testbed dataset is elaborated in the following section. 

\subsection{Server Configurations and Users Specifications}
\label{config}
The set up process of the FreeRADIUS server is performed on a Linux machine with 2.30 GHz Intel(R) Core(TM) i5-2410M CPU, and 8GiB System Memory. The database system used to store primary configurations and AAA information is MySQL and consist of 10 preordained tables. The principal tables employed for data collection and analysis are labeled as radcheck (authentication), radpostauth (authorization) and radacct (accounting). Other essential configurations are conducted directly on FreeRADIUS setting files, such as server and client security configurations, required certificates, database setups, and so forth. 

As stated earlier, the testbed deployed for this study is dedicated to one-AP-many-users. Thus, we describe the AP configurations and wireless users specifications in the following lines. The AP is an enhanced 802.11g wireless access point powered by D-Link 108G technology, DWL-2100AP, and supports WPA and WPA2 security protocols. The wireless users connected to this network during one month of experiment consist of two laptops, two smart phones, and two tablets. A summary of the users' specifications in terms of devices, operating systems and participation time in the experiment is provided in Table \ref{tab:4}. Obviously not all the users were present everyday and every hour of the test, but they follow a natural form of entering and exiting the network. Some devices were disassociated from the network when the users simply depart from the coverage area and others were deliberately disconnected in the time of specific anomaly generation. In the coming sections we present all types of anomalies generated and organized for this testbed.

\begin{table}[!htbp]
\centering
\caption{A summary of the testbed users' specifications}
\label{tab:4}       
\begin{tabular}{| p{3.1cm} | p{1.8cm} | p{1.8cm} |}
\hline
Device & OS & Participation Time (\%)  \\
\hline\hline
Surface Pro II & Win 10 & 100\% \\ \hline
Asus & Win XP & 100\% \\ \hline
Alcatel onetouch & Andriod & 85\% \\ \hline
iPhone & iOS & 15\% \\ \hline
iPad & iOS & 100\% \\ \hline
Dell & Win 10 & 8\% \\ \hline
\end{tabular}
\end{table}

\subsection{Network Anomaly Generation in a Controlled Environment}
\label{anom-gen}
In this section we describe how some of the known wireless network issues are re-generated to make the desired data records for the evaluation of the proposed methods in this work. In the time of experiment not all days encounters anomalies, some days simply end as NORMAL days and the users' connection and amount of network usage are according to the users' usual plan of the day. In ABNORMAL days, however, one or some kind of anomalies are provoked to test the behavior of the model under abnormal circumstances. The anomalous patterns selected for this purpose are common cases occur in real networks relatively often and affect the performance of users connection and availability of the network. Succeeding paragraphs deal with the specific aspects of these anomalies and point out how to replicate them. 

\subsubsection{AP Shutdown/Halt}
\label{ap-halt}
To reproduce this anomalous effect, when there is no session recorded in the accounting table, the AP could be shutdown for a while or restarted. This anomaly is regenerated under various circumstances and for different period of time and in the real world could be considered as AP shutdown, halt, crash or power off. 

\subsubsection{Heavy Usage}
\label{heavy-usage}

\paragraph{Single User}
\label{heavy-usage-one}
This anomaly arises when only one user performs heavy download or upload. It might affect the rest of the associated users depending on the amount of usage, duration, time of the day and other relevant factors.

\paragraph{Multiple Users}
\label{heavy-usage-many}
This anomaly emerges when more than one user use the network excessively, and therefore the overall throughput of the network intensifies. This could occur in a NORMAL day or as an anomalous event and the network tolerance, as expected, varies for different networks and different AP configurations. In any case, the proposed model is expected to detect the irregularity and report the level of hazard so that the network managers could take control of the situation and make required changes if possible.     

\subsubsection{Wireless Network Interference}
\label{interfer}
In a real network, a variety of things can interfere with the radio waves, degrading the quality of connection and decreasing the network reliability. Sources of interference are commonly from other wireless networks in the vicinity when they all locate in the same channel, from non-802.11 devices such as microwave ovens or cordless phones that use 2.4GHz band as well, from other clients in a crowded environment when they all try to transfer data at the same time, and from RF effects such as hidden terminals or capture effects. In this work we intend to cause interference anomaly in a systematic and controlled manner. For this aim, we made use of a python script named  wifijammer \cite{Ref29} to intentionally jam wireless clients or APs in the range to simulate the same outcome as the aforementioned interferences. The jamming process works by sending 1 de-authentication packet to the client from the AP, 1 de-auth to the AP from the client, and 1 de-auth to the AP destined for the broadcast address to de-authenticate all clients connected to the AP. Many APs, however, ignore de-auth to broadcast addresses.  
We employed wifijammer in the following plans by applying peculiar properties each time to create different forms of interferences. 

\paragraph{Jamming the Entire Channel}
In this practice, the monitor mode interface is set to listen and de-authenticate clients or APs on a specific channel. This way of jamming influence all the available networks on the current channel and imply interferences caused by busy channels. 
     
\paragraph{Jamming Clients with Various Time Intervals}
Executing the De-authentication procedure with short time intervals hinder clients from recovering and disable them for the entire period of jamming, so the immediate result in the accounting table is the one-time stop session from each client and then a silent period without any start session. While de-authenticating with a larger time interval makes clients reclaim and try to get back the connection to the AP, and subsequently many short sessions is observed in the accounting table because they are de-authenticated right after getting connected again. In such manner we can replicate two interference cases observed in the real datasets frequently. 

\paragraph{Jamming Specific Clients}
De-authenticating some specific clients and not the rest, resembles the hidden-terminal situation, when one client is forced to back-off and delay data transfer because the other clients can not sense its send-request. Depending on the time interval discussed earlier, the sessions outcome in the accounting table could be different.

\subsection{Testbed Experimental Results}
\label{testbed-exp-res}
The testbed experiment is deployed in a home environment, with a single AP and 6 regular users and between 3-4 guest users. The experiment contains 5 weeks of data, 30 working days, and is performed in two different time span, once in November 2015 and a while later in April 2016. There exist 20 normal days with no anomalies provoked, and 10 abnormal days containing at least one anomalous event a day. Each anomaly takes from 15 minutes to around an hour. 

In the following paragraphs we show how the modeling and anomaly detection techniques operate in the presence of the ground truth, data obtained from the testbed deployment.

\subsubsection{GMM vs. HMM Modeling: Pros and Cons}
\begin{figure}
\centering
\includegraphics[width=0.5\textwidth]{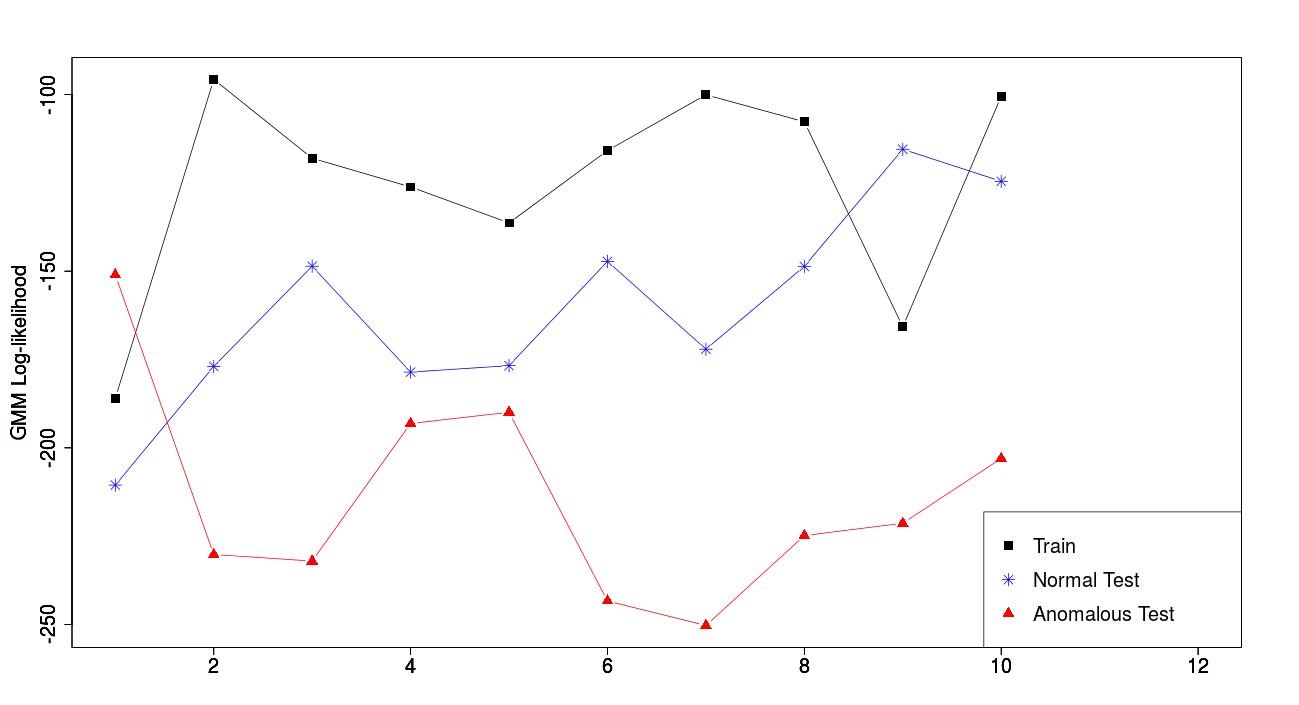}
\caption{Likelihood values of the training and test data belong to Testbed for GMM Model}
\label{fig:50}      
\end{figure}

\begin{figure}
\centering
\includegraphics[width=0.5\textwidth]{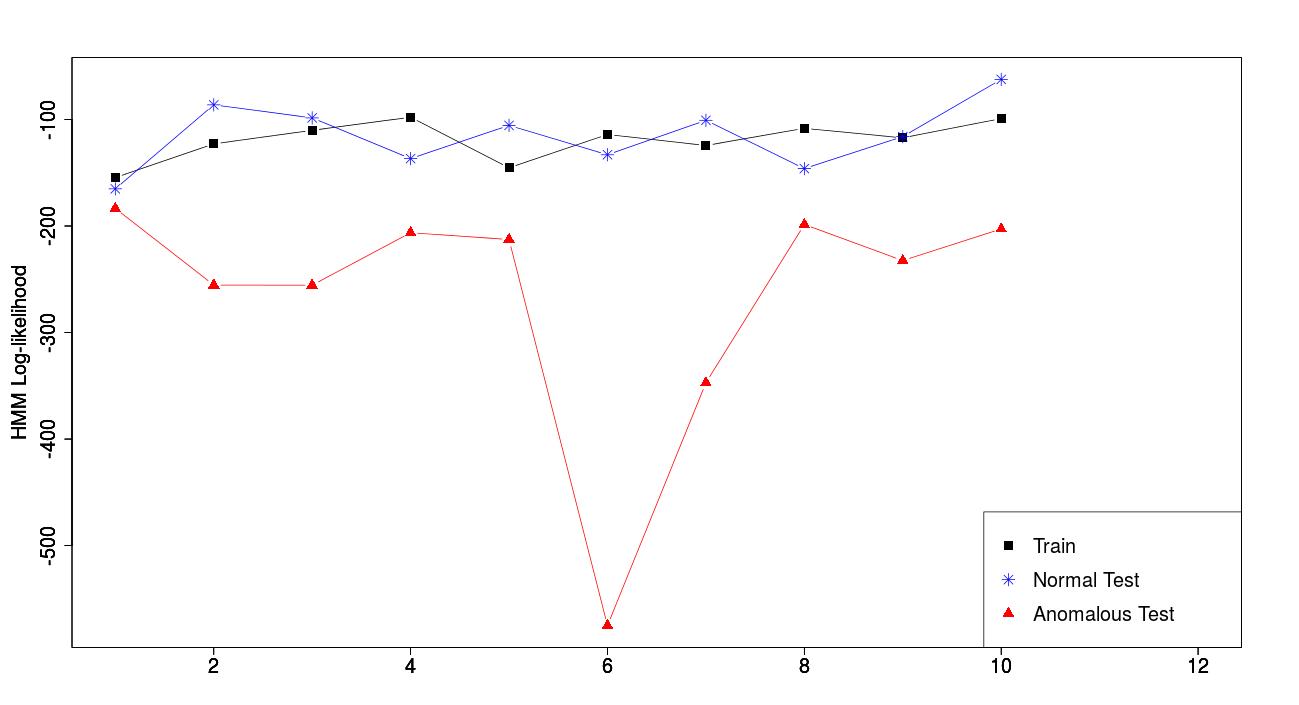}
\caption{Likelihood values of the training and test data belong to Testbed for HMM model}
\label{fig:51}      
\end{figure}

For the first experiment, a GMM model is built with 10 randomly selected normal days as training data. From then on, the likelihood of the generated model is computed against the training data as well as 10 unobserved normal days and 10 abnormal days as test data. The same process is performed on the HMM model, with the same set of training and test data. The summary of this experiment is displayed in Figure \ref{fig:50} and \ref{fig:51}. 

\noindent

\noindent

\begin{table*}[!htbp]
\centering
\caption{Anomaly detection of the normal and anomalous test data belong to Testbed for GMM}
\label{tab:5}
\begin{tabular}{|l|p{2.6cm}|p{2.7cm}|p{2.6cm}|p{1.25cm}|p{1.2cm}|}\hline
\diaghead(-5,1){\hskip5.0cm}%
    {Data - Threshold}{Statistical Metrics}&
False Positive Rate (FPR) & True Negative Rate (TNR) & True Positive Rate (TPR) & Accuracy (ACC) & F1 Score\\
\hline \hline
Normal Testset (Threshold: 0.6) & 2.5\% & 97.5\% & 0\% & 97.5\% & 0\% \\ \hline
Normal Testset (Threshold: 0.7) & 5.5\% & 94.5\% & 0\% & 94.5\% & 0\% \\ \hline
Normal Testset (Threshold: 0.8) & 10.5\% & 89.5\% & 0\% & 89.5\% & 0\%  \\ \hline \hline
Anomalous Testset (Threshold: 0.6) & 3\% & 97\% & 4.7\% & 81\% & 8.1\% \\ \hline
Anomalous Testset (Threshold: 0.7) & 9.9\% & 1.01\% & 4.7\% & 75\% & 7.2\%  \\ \hline
Anomalous Testset (Threshold: 0.8) & 19\% & 81\% & 24.9\% & 70\% & 20.75\%  \\ \hline
\end{tabular}\medskip
\end{table*}

\begin{table*}[!htbp]
\centering
\caption{Anomaly detection of the normal and anomalous test data belong to Testbed for HMM}
\label{tab:6}
\begin{tabular}{|l|p{2.6cm}|p{2.7cm}|p{2.6cm}|p{1.25cm}|p{1.2cm}|}\hline
\diaghead(-5,1){\hskip5.0cm}%
    {Data - Threshold}{Statistical Metrics}&
False Positive Rate (FPR) & True Negative Rate (TNR) & True Positive Rate (TPR) & Accuracy (ACC) & F1 Score\\
\hline \hline
Normal Testset (Threshold: -50) & 0.5\% & 99.5\% & 0\% & 99.5\% & 0\% \\ \hline
Normal Testset (Threshold: -20) & 1.75\% & 98.25\% & 0\% & 98\% & 0\% \\ \hline
Normal Testset (Threshold: -10) & 3.75\% & 96.25\% & 0\% & 96\% & 0\%  \\ \hline \hline
Anomalous Testset (Threshold: -50) & 0\% & 100\% & 39\% & 90\% & 49\% \\ \hline
Anomalous Testset (Threshold: -20) & 0\% & 100\% & 43\% & 91\% & 52\%  \\ \hline
Anomalous Testset (Threshold: -10) & 1.1\% & 98.9\% & 75\% & 95\% & 74\%  \\ \hline
\end{tabular}\medskip
\end{table*}

\begin{table*}[!htbp]
\centering
\caption{Detection rate of various anomalous patterns of the Testbed}
\label{tab:7}
\begin{tabular}{|l|p{2.5cm}|p{2.5cm}|p{1.8cm}|p{1.95cm}|p{1.6cm}|}\hline
\diaghead(-5,1){\hskip5.0cm}%
    {Model}{Anomalous Patterns}&
Jamming Channel (Low Intervals) & Jamming Channel (High Intervals) & Heavy Usage (Single User) & Heavy Usage (Multiple Users) & AP Power Off\\
\hline \hline
GMM (Threshold: 0.8) & 28.5\% (4/14) & 17.3\% (4/23) & 8.3\% (1/12) & 0\% (0/3) & 35.2\% (6/17)  \\ \hline \hline
HMM (Threshold: -10) & 71.4\% (10/14)& 73.9\% (17/23)& 83.3\% (10/12)& 100\% (3/3)& 82.3\% (14/17) \\ \hline
\end{tabular}\medskip
\end{table*}

Both figures demonstrate overall higher likelihood values for the training data. The likelihood values of the unobserved data set is divided into normal and abnormal outputs which are displayed in graphs with different colors and shapes. In both models there are higher likelihood values for the normal days rather than the abnormal days. However, there is a discernible boundary between the normal and abnormal results in HMM while in GMM the likelihood values are not clearly separated and there are even some instances that the likelihood value of the normal day is lower than the abnormal day. The daily likelihood of abnormal days are apparently lower than the normal days, and this value varies with the number of abnormal occurrences and duration of each event. However, it is more straightforward to define a threshold for HMM rather than GMM model, to announce a day normal or abnormal.

\subsubsection{Anomaly Detection}
In this section we determine the anomalous time-slots with the proposed methodologies and compare the achieved results from the model with the testbed anomalous ranges recorded for the abnormal instances. Note that various thresholds for each technique produce different results as the detection and false positive rates change based on the selected threshold. We made use of some statistical metrics to measure the detection accuracy and false alarms such as fall-out or false positive rate (FPR), specificity (SPC) or true negative rate (TNR), sensitivity or true positive rate (TPR), and eventually accuracy (ACC) and F1 score. In order to acquire the specific definition of each terminology refer to \cite{Ref36}.

The summary of the analysis on the normal and anomalous test data are presented in Table \ref{tab:5} for GMM modeling and in Table \ref{tab:6} for HMM modeling approaches.

Table \ref{tab:5} shows that higher thresholds increase the possibility of anomaly detection (24.9\% rather than 4.7\%), however the false positive rates also increase accordingly (19\% rather than 9.9\% and 3\%). In normal test data, when we expect no anomalies to occur, from 2.5\% to 10.5\% fall-out is observed. Comparing this fall-out ratio to the results of Table \ref{tab:6} for normal test set, it is noted that much lower false alarms is marked for HMM (from 0.5\% to 3.75\%). Furthermore, the FPR for the anomalous data in HMM is quite trivial relative to GMM FPR output (1.1\% in HMM vs. 19\% in GMM). The highest detection rate or TPR in HMM modeling is achieved with Threshold equals to -10 which is 75\% in average for 10 abnormal days of the experiment.  

Regarding the FPR or fall-out ratio recorded for normal data in HMM, a careful consideration on each false alarm is performed and it is noted that the HMM model is slightly sensitive to extreme download ratio and in some cases both download and upload volumes. As the testbed is deployed in a real home environment with real wireless users, although in normal days no anomaly is generated deliberately, there might have been some evidences of rather high download or upload by the users as it happens quite often in every wireless network. Therefore the false positive examples occurred in normal days could be introduced as real anomalies appearing in normal days, however for this experiment we assumed that normal days contain no anomalies. In our future work we intend to propose an unsupervised anomaly detection algorithm that detect anomalies in an unlabeled test dataset which is the case when no ground truth is actually provided.

Table \ref{tab:7} displays the total proportion of different anomalies' occurrences in the Testbed and presents the detection rate of each anomalous pattern by GMM and HMM. Here we consider the anomalous test data and the highest likelihood thresholds of both models (0.8 for GMM and -10 for HMM) that provide the maximal detection rate. Detection ratio is determined by the overall number of time-slots marked as anomaly by the model divided by the total number of time-slots encounter particular types of anomaly. 
Comparing GMM and HMM once more demonstrates the superior capability of HMM in recognition of anomalous events, while providing unnoticeable false positive rate (Table \ref{tab:6}). Among the various types of anomalies generated for the Testbed, the highest detection rate belongs to heavy usage pattern, producing by multiple users and then single user. The lowest detection ratio, however, originates from jamming channel with low interval. Although there are some specific anomalous instances that are never detected by the model, regardless of the cut-off threshold, the overall detection rate of the HMM is quite satisfactory. We intend to improve the detection estimate capacity of HMM in our future work by proposing more complex variations of HMMs. 

\subsection{Conclusions}
In this section we described the Testbed deployment on a single AP and the process of data collection from a RADIUS server. We further explained the anomalies generated deliberately to prepare the ground truth data for model evaluation. We reproduced AP Shutdown/Halt, Heavy Usage from a single user and multiple users, and various types of interferences as a set of network anomalies. We then applied our proposed model to detect anomalous points, and discussed the effectiveness of each model. The experimental results demonstrated that HMM outperformed GMM in obtaining higher detection ratio while producing minor false alarm.  

\section{Conclusions and Future Work}
In large deployments of 802.11 networks with varying usage, channel conditions, and operational constraints,  network managers often demand tools that provide them with a comprehensive view of the entire network. Analyzing the users' behavioral patterns, learning efficient models to detect anomalous periods, and measuring the temporal performance of the network under certain circumstances are of great significance to provide an adequate level of satisfaction for the wireless users. Proposing time-invariant and time-variant modeling approaches and utilizing those models for anomaly detection in addition to a RADIUS testbed deployment with simulated anomalies compose the key contributions of this work.

We proposed a new application of HMMs in performance anomaly detection of 802.11 wireless networks and explored the necessity of temporal specifications of HMM rather than its simple time-independent counterpart model, GMM. We performed analysis and compared HMM and GMM in terms of modeling competency and anomaly detection performance on the large FEUP dataset as well as a similar but minor version of the deployed testbed with provoked anomalies for evaluation purposes. 

The experimental results show that HMM models are capable of detecting a great portion of provoked anomalies on unobserved test data set (up to 75\% TPR), and even disclosing unintentional anomalies occurred during the normal days of experiment. Besides, the false positive ratio is fairly low (only 1.1\%) in HMM that outperforms GMM both in detection and fall-out rate. 
 
In future work we intend to propose an anomaly detection algorithm that works in unsupervised mode regardless of the anomalous information provided for the data records. Furthermore, we will propose more complex HMMs to characterize and distinguish various anomaly-related patterns. We also plan to extend the testbed to multiple APs to explore new aspects of anomalies that concern the mobility effects of the wireless users in AP vicinities.     

\begin{acknowledgements}
This work is financed by the ERDF – European Regional Development Fund through the Operational Programme for Competitiveness and Internationalisation - COMPETE 2020 Programme within project «POCI-01-0145-FEDER-006961», and by National Funds through the FCT – Fundação para a Ciência e a Tecnologia (Portuguese Foundation for Science and Technology) as part of project UID/\\EEA/50014/2013. The first author is also sponsored by FCT grant SFRH/BD/99714/2014.
\end{acknowledgements}

\bibliographystyle{plain}
\bibliography{mybiblo.bib}   
  
\newpage

\parpic{\includegraphics[width=1in,clip,keepaspectratio]{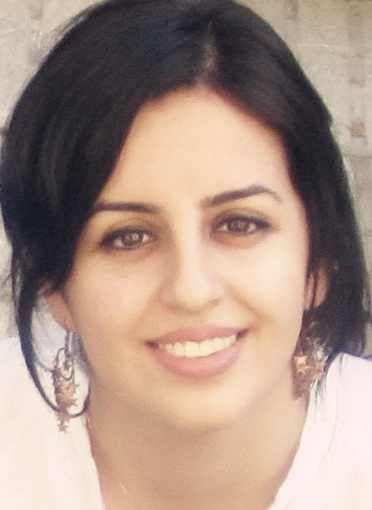}}
\noindent 
\begin{small}
{{\bf Anisa Allahdadi} received the B.Sc. in Computer Science from BIHE University (Bah\'{a}'\'{i} Institute for Higher Education), Iran in 2006 and M.Sc in Software Engineering from BIHE University, Iran in 2010. 
She is currently a researcher in the Center for Telecommunications and Multimedia at INESC TEC and pursuing her Ph.D in the MAP-i Doctoral Programme in the Faculty of Engineering of University of Porto. Her research interest include network management, probabilistic modeling and anomaly detection in IEEE 802.11 based wireless networks.}
\end{small}

\vspace{5 mm}
\parpic{\includegraphics[width=1in,clip,keepaspectratio]{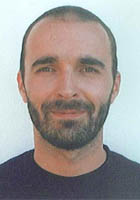}}
\noindent 
\begin{small}
{{\bf Ricardo Morla} 
is an assistant professor of electrical and computer engineering at the University of Porto and principal investigator at INESC Porto, Portugal. His research interests are in the area of automatic system management with an emphasis on probabilistic modeling, prediction, anomaly detection, and root-cause analysis for ICT systems including network infrastructure and smart environments. He holds a Ph.D. in computer science from
Lancaster University UK.}
\end{small}
\end{document}